\documentclass[%
 reprint,
superscriptaddress,
nofootinbib,
 amsmath,amssymb,
 aps,
floatfix,
prd
]{revtex4-2}

\usepackage{graphicx}
\usepackage{dcolumn}
\usepackage{bm}
\usepackage{float}
\usepackage[usenames,dvipsnames]{color}
\usepackage{caption}
\usepackage{subcaption}
\usepackage{comment}
\usepackage[table]{xcolor}
\usepackage{booktabs}
\usepackage{braket}
\usepackage{mathtools}
\usepackage[normalem]{ulem}

\begin{document}
\newcommand{\bea}{\begin{eqnarray}}
\newcommand{\eea}{\end{eqnarray}}
\newcommand{\beq}{\begin{equation}}
\newcommand{\eeq}{\end{equation}}
\newcommand{\sean}[1]{\textcolor{red}{\textbf{Sean: #1}}}
\newcommand{\anuj}[1]{\textcolor{blue}{ #1}}

\def\psi4{$\Psi_4$}
\def\news{$\mathcal{N}$}
\def\strain{$h$}
\def\t0{$t_0$}
\def\leq{\,\raise 0.4ex\hbox{$<$}\kern -0.8em\lower 0.62ex\hbox{$-$}\,}
\def\geq{\,\raise 0.4ex\hbox{$>$}\kern -0.7em\lower 0.62ex\hbox{$-$}\,}
\def\lsim{\,\raise 0.4ex\hbox{$<$}\kern -0.75em\lower 0.65ex\hbox{$\sim$}\,}
\def\gsim{\,\raise 0.4ex\hbox{$>$}\kern -0.75em\lower 0.65ex\hbox{$\sim$}\,}
\def\pm{\,\raise 0.4ex\hbox{$+$}\kern -0.75em\lower 0.65ex\hbox{$-$}\,}

\title{BOB the (Waveform) Builder: Analyzing Analytical Black-Hole Binary Merger Waveforms with the Backwards One Body Formalism}

\author{Anuj Kankani}
\email{anuj.kankani@mail.wvu.edu}
\affiliation{Department of Physics and Astronomy, West Virginia University, Morgantown, WV 26506, USA
}
\affiliation{Center for Gravitational Waves and Cosmology, West Virginia University, Chestnut Ridge Research Building, Morgantown, WV 26505,
USA
}
\author{Sean T. McWilliams}
\affiliation{Department of Physics and Astronomy, West Virginia University, Morgantown, WV 26506, USA
}
\affiliation{Center for Gravitational Waves and Cosmology, West Virginia University, Chestnut Ridge Research Building, Morgantown, WV 26505,
USA
}

\date{\today}

\begin{abstract}
The Backwards-One-Body (BOB) model provides a fully analytical and physically motivated description of the merger-ringdown gravitational radiation emanating from a black hole binary merger. We perform a comprehensive validation of BOB for the dominant $(2,2)$ mode of quasi-circular and non-precessing systems, assessing its accuracy against numerical relativity (NR) simulations, state-of-the-art waveform models, and a sum of quasinormal modes. We demonstrate that BOB most accurately describes the gravitational wave news, achieving accuracy comparable to highly-calibrated Effective-One-Body and NR surrogate models. Because BOB is minimally tuned to NR catalogs, it retains a high level of accuracy in regions of the parameter space sparsely covered by current NR catalogs. BOB yields an analytic link between the amplitude of the fundamental quasinormal mode and the peak amplitude of the news, which we verify to within the errors of a surrogate ringdown model. We identify a flavor of BOB that requires only the remnant mass and spin, yet matches the accuracy of models that fit a sum of many overtones. Lastly, we show that BOB accurately models both the mass and current quadrupole waves for superkick configurations, contrary to a claim in the literature, and explain why that study was not actually implementing BOB as it has been defined. Our findings establish BOB as a powerful tool for gravitational wave analysis, for providing independent tests of NR-calibrated models, and for better understanding the underlying physics of the merger. We provide a companion python package, \texttt{gwBOB}, allowing for the easy construction of various flavors of BOB and comparison to NR waveforms.
\end{abstract}

\maketitle


\section{Introduction}
In order to realize the full potential of gravitational wave astronomy, accurate and efficient gravitational waveform models are of critical importance. Although the most accurate waveforms are produced through numerical relativity (NR) simulations, the significant computational expense of these simulations makes it impossible to directly rely on NR to generate the enormous number of waveform realizations required for parameter estimation. Instead, analytic approximations must be combined with NR information to provide the most accurate and computationally efficient gravitational waveform models. Multiple semi-analytic waveforms have been developed to tackle this problem, encompassing a variety of methods. NR surrogate \cite{surr_fast,surr_cite2,surr_cite1,surr_cite3,surr_cite4,surr_cite5,cce_surr,surr_ringdown} based waveforms take a data driven approach by directly interpolating NR waveforms, but benefit from hybridization with post-Newtonian (PN) \cite{PN_cite1,PN_cite2} inspiral waveforms for longer inspirals \cite{surr_cite1}. Effective-One-Body \cite{EOB_cite1,EOB_cite2,SEOBNR_cite1,TEOB_cite1,gamba2022effective,SEOBNR_cite2,SEOBNR_cite3,nagar2023analytic} methods combine analytical results from PN, resummation techniques, e.g. \cite{ damour1998improved, damour_resummation}, with an ad hoc merger-ringdown ansatz, as well as extensive calibration of the model to NR simulations, including fixing free higher-order parameters in the inspiral. IMRPhenom based approaches \cite{phenom_cite1,phenom_cite2,phenom_cite3,phenom_cite4,phenom_cite5,phenom_cite6,hamilton2025phenomxpnr} combine analytical and NR information to produce computationally efficient waveform models, ideal for data analysis pipelines.

The increasing sensitivity of current and future ground and space-based detectors \cite{abac2025gwtc,abbott2020prospects,LVK_ObservingPlans_2025,LISA,CE,Einstein_tel} demands a corresponding increase in waveform template accuracy \cite{purrer2020gravitational,ferguson2021assessing,hu2022assessing,jan2024accuracy}. Due to the large computational cost of NR simulations, gravitational waveform catalogs \cite{sxs_cat1,sxs_cat2,sxs_cat3,cce_website,rit_cat2,rit_cat3,rit_cat4,maya_cat1,maya_cat2,hamilton2024catalog,cce_website,trenado2025first_catalog} will only have limited coverage over the parameter space, which will limit the accuracy of heavily calibrated semi-analytic models. In this work we show how the accuracy of these models drops in sparsely covered regions, while the accuracy of the minimally tuned BOB remains robust.

The Backwards-One-Body (BOB) model \cite{BOB} is an analytical and physically motivated approach to modeling the merger-ringdown radiation emitted from a black hole binary merger, based on the behavior of a bundle of null geodesics diverging from the light ring of the final remnant black hole spacetime. While prior work has explored BOB in specific contexts, such as for superkick configurations \cite{universal_superkick}, as a replacement for NR-calibrated merger-ringdowns in EOB based inspiral-merger-ringdown (IMR) waveform models \cite{Mahesh:2025}, and for predicting the peak times of higher order modes \cite{Kankani_HM_tp}, a systematic analysis of its accuracy across the parameter space has been missing. Given that the BOB equations permit several ``flavors'' based on the approach to choosing initial conditions and the choice of which gravitational wave quantity to directly model (i.e., strain or its first or second time derivative), it is critical to gain a better understanding of how to maximize the accuracy of BOB across the entirety of the physical parameter space, without sacrificing BOB's physics-first approach by introducing ad hoc calibration to NR. In this work, we focus on the dominant $(2,2)$ mode of quasicircular and non-precessing configurations and conduct a comprehensive assessment of the BOB's accuracy, thereby determining the methods that maximize the accuracy of BOB while still retaining its physical approach to waveform modeling. It is important to note that while we focus on the $(2,2)$ mode for non-precessing systems in this work, most IMR waveform models used in data analysis pipelines are capable of modeling precessing systems and higher modes \cite{abac2025gwtc,abac2025gwtc_methods}.

We show that BOB's physical foundation and high accuracy near the merger provide a powerful tool for gaining new insights into the strong field dynamics of black hole binary mergers. In \cite{McWilliams_SK:2025}, we demonstrate that the broader paradigm that BOB first introduced, i.~e.~the notion that the spacetime will behave like a perturbed version of the remnant spacetime throughout the merger, can predict in detail the behavior of spin-induced recoils, including the so-called ``superkick'' configurations; this behavior is a manifestly strong-field phenomenon.  In this work, in addition to establishing BOB's accuracy, we explore the similarities between BOB and overtone-based models. We provide several comparisons that demonstrate how BOB's analytic modeling of the merger can be used to illuminate the underlying physics of black hole binary mergers and the connection between the merger and the ringdown radiation.

For many scenarios, the accuracy of BOB can be increased by introducing additional NR information, depending on the problem being addressed. For example, as done in \cite{Mahesh:2025}, to enforce continuity with an EOB inspiral, information about the amplitude and frequency at the time of the peak strain amplitude was utilized. Similarly in \cite{Kankani_HM_tp}, peak frequency information was combined with BOB to predict the time at which subdominant gravitational wave modes peak. However, in this work we will focus on assessing the accuracy of BOB as a standalone merger-ringdown waveform and explore which flavor of BOB provides the highest accuracy while requiring the least amount of tuning to NR.  In particular, we introduce and validate a flavor of BOB that requires only the remnant mass and spin as inputs, yet achieves an accuracy comparable to state-of-the-art models that rely on significantly more NR information. We also provide a python package, \texttt{gwBOB} \cite{gwBOB}, simplifying the process for constructing various flavors of BOB and comparing them to NR waveforms. 

This paper is structured in the following way: In Sec.~\ref{sec:constructing_BOB} we explain how we construct BOB, and show that BOB does not naturally model the strain, but rather the strain must be obtained by integrating BOB after building it for the news, \news, or the Weyl scalar $\Psi_4$. In Sec.~\ref{sec:results} we compare BOB to a sum of quasinormal modes, including assessing BOB's accuracy at predicting the amplitude of the fundamental mode, as well as the merger ringdown portion of IMR waveform models. In Sec.~\ref{sec:superkicks} we show that BOB, contrary to what was shown in \cite{universal_superkick}, does not exhibit a bias for the current quadrupole wave. Finally, in Sec.~\ref{sec: discussion} we summarize our results and discuss the unique benefits the BOB approach to merger ringdown modeling can provide to the waveform modeling community. Throughout this paper we set $G=c=1$ and $M$ refers to the sum of the Christodoulou masses of the individual horizons defined at \texttt{reference\_time} in the SXS metadata \cite{sxs_cat3}. For all NR simulations considered in this work, the default extrapolation order is used and we use the v3.0.0 SXS catalog. The selection of quasicircular and non-precessing simulations is obtained through the default selection criteria available through the SXS catalog. The dataset spans nearly 600 simulations, up to mass ratio $q$ of 20 and initial spin magnitudes of $|\vec{\chi}_i|\approx 0.99$. Throughout this work, we do not include the mismatch between the highest and second-highest resolution NR waveform in our plots, but do include differences when showing waveforms from specific simulations. The median mismatch for the NR news between the highest and second highest resolution levels, from the peak to $75M$ afterwards, is $2.83 \times 10^{-8}$, which is well below the median mismatch between NR and waveform models. Thus the inherent errors in NR is not a limiting factor in this analysis. We note there are a very small number of individual cases in which different waveform models may have lower mismatch compared to NR than the NR mismatch between resolution levels. There are 3 such cases for BOB, 5 cases for SEOBNRv5 and 12 for NRHybSur3dq8. The NR news used in this work is obtained through a cubic spline differentiation of the NR strain. When CCE waveforms \cite{cce_cat,cce_citation} are considered, we use the second smallest worldtube radius as suggested by \cite{cce_website}.

\section{Constructing BOB: CAN WE BUILD IT? YES WE CAN!}
\label{sec:constructing_BOB}
BOB models the merger and ringdown radiation by considering the motion of a bundle of null geodesics, i.~e.~ a null congruence, centered on the light ring of the remnant black hole. Using the geometric optics approximation, the transport equation can be used to connect the cross-sectional area of the null congruence to a waveform amplitude \cite{BOB,qnm_lightring1,qnm_lightring2} given by
\begin{equation}
    A = A_p\, \text{sech}\bigg(\frac{t-t_p}{\tau}\bigg),
    \label{eq:BOB_amplitude}
\end{equation}
where $A_p$ is the peak amplitude of the waveform, $t_p$ is the time of the peak amplitude, and $\tau$ is the damping time associated with the fundamental quasinormal mode.
In principle the underlying amplitude evolution of BOB can be used to model either \strain~ or any of its derivatives. In \cite{BOB}, one of us (STM) suggested that at early times this amplitude evolution might most accurately describe \psi4, based on the behavior of perturbations around a slowly spinning black hole. However, \cite{universal_superkick} showed that for superkick configurations, the BOB amplitude evolution best described the news \news. Depending on the gravitational quantity we choose to model using Eq.~\eqref{eq:BOB_amplitude}, the BOB frequency evolution will also change. In \cite{BOB} the news and frequency are connected through the relation
\begin{equation}
    |R\mathcal{N}_{lm}|^2 = 16\pi m^2 \frac{dJ^{\text{orb}}_\text{lm}}{d\Omega_{lm}}\Omega_{lm}\dot{\Omega}_{lm} \label{eq:News_frequency},
\end{equation}
where $\Omega_{lm}$ is the implied orbital frequency of the radiation source, $m$ is the gravitational wave mode number, and the orbital angular momentum spectral density $\frac{dJ^\text{orb}_{lm}}{d\Omega_{lm}}$ is a constant to excellent approximation throughout the late inspiral and merger-ringdown, as shown in \cite{baker2008mergers}.\footnote{For the rest of our equations we suppress the $(l,m)$ subscript in $\mathcal{N}_{lm}$ and the factor of the radius $R$.}
In \cite{BOB}, Eq.~\eqref{eq:BOB_amplitude} was taken to describe $\Psi_4$ which can be related to \news~ through
\begin{align}
     \Psi_4 &= \frac{d}{dt}\big(|\mathcal{N}|e^{-i\phi}\big)\\  
     &= \left(-i\omega|\mathcal{N}| + \frac{d}{dt}|\mathcal{N}|\right) e^{-i\phi}\nonumber,
\end{align}
where $\omega \equiv \frac{d\phi}{dt}$ is the gravitational wave frequency.
With the assumption that near merger $\omega|\mathcal{N}| \gg \frac{d}{dt}|\mathcal{N}|$, the above equation simplifies to
\begin{equation}
    |\Psi_4| \approx \omega |\mathcal{N}| \label{eq:psi4_news}.
\end{equation}
The waveform frequency $\omega$ is related to the orbital frequency $\Omega$ by the relation $\omega = m\Omega$, where $m$ is the gravitational wave mode number. 
Eq.~\eqref{eq:psi4_news} then allows us to solve Eq.~\eqref{eq:News_frequency} under the assumption that Eq.~\eqref{eq:BOB_amplitude} best describes \psi4. The full frequency and phase terms for $\Psi_4$ can be found in \cite{BOB}.
Following the same line of reasoning, we can then relate \strain\, to \news\, by 
\begin{equation}
    |h| \approx \frac{|\mathcal{N}|}{\omega} \label{eq:strain_news},
\end{equation}
and obtain a frequency under the assumption that Eq.~\eqref{eq:BOB_amplitude} best describes the strain \strain. The resulting frequency is given by
\begin{equation}
    \Omega = \Omega_{\text{QNM}}\bigg(\frac{\Omega_0}{\Omega_\text{QNM}}\bigg)^G,
\end{equation}
where 
\begin{equation}
    G \equiv \frac{\tanh\left(\frac{t-t_p}{\tau}\right)-1}{\tanh\left(\frac{t_0-t_p}{\tau}\right)-1}.
\end{equation}
The phase is then given by
\begin{equation}
\begin{split}
    \Phi = \frac{\tau\Omega_{\text{QNM}}}{2}\bigg[\bigg(\frac{\Omega_0}{\Omega_{\text{QNM}}}\bigg)^K\text{Ei}(B_+) - \text{Ei}(B_-)\bigg] + \Phi_0.
\end{split}
\end{equation}
where 
\begin{equation}
\begin{split}
    K &= \frac{2}{\tanh\bigg(\frac{t_p - t_0}{\tau}\bigg) + 1},\\
    B_{\pm} &=\frac{-\ln\bigg(\frac{\Omega_0}{\Omega_\text{QNM}}\bigg)\bigg(\tanh\bigg(\frac{t - t_p}{\tau}\bigg) \pm 1\bigg)}{\tanh\bigg(\frac{t_0 - t_p}{\tau}\bigg)-1},
\end{split}
\end{equation}
and Ei is the exponential integral \cite{Ei}. Here, the frequency $\Omega$ is related to the waveform frequency, $\omega$, through $\omega = m\Omega$ and $\Phi$ is related to the waveform phase, $\phi$, through $\phi = m\Phi$. Throughout this paper, we will take $m=2$ due to our focus on the $(2,2)$ mode. $\Omega_\text{QNM}$ is therefore related to the real frequency of the fundamental quasinormal mode $w_{220}$ as $\Re(\omega_{220}) = 2\Omega_\text{QNM}$. $t_0$ and $\Omega_0$ represent initial condition parameters, where $\Omega_0$ is the frequency reached at $t_0$. $\Phi_0$ represents an arbitrary phase shift. Unlike in \cite{BOB, Mahesh:2025}, where $t_0-t_p$ was constrained to ensure continuity with an inspiral, here we treat the initial conditions with more flexibility. Since in this work we focus only on the merger-ringdown radiation, we find that taking the limit, $t_0 = -\infty$, provides a useful simplification of our expressions without sacrificing accuracy, provided $\Omega_0$ is chosen according to the procedure described in Sec.~\ref{sec:IC}.

Alternatively, we can assume that the amplitude evolution of Eq.~\eqref{eq:BOB_amplitude} best describes \news. This scenario allows us to directly solve Eq.~\eqref{eq:News_frequency} without needing to assume adiabaticity. The resulting frequency is then given by
\begin{align}
    \Omega = \sqrt{\Omega^2_\text{{QNM}} + F\bigg[\tanh\bigg(\frac{t-t_p}{\tau}\bigg) - 1\bigg]},
    \label{eq:BOB_news_frequency_evolution}
\end{align}
where
\begin{align}
    F \equiv \frac{\Omega^2_\text{QNM}-\Omega^2_0}{1-\tanh\left(\frac{t_0-t_p}{\tau}\right)}.
\end{align}
For finite $t_0$, under the condition that $F - \Omega_{\text{QNM}}^2>0$ the phase has the closed form expression
\begin{equation}
\begin{split}
    \Phi &= \frac{\tau\Omega_{\text{QNM}}}{2}\ln\bigg(\frac{\Omega + \Omega_{\text{QNM}}}{|\Omega - \Omega_{\text{QNM}}|}\bigg) \\&- \tau\sqrt{2F - \Omega_{\text{QNM}}^2}\arctan\bigg(\frac{\Omega}{\sqrt{2F - \Omega_{\text{QNM}}^2}}\bigg) + \Phi_0.
\end{split}
\label{eq:News_phase_finite_t0}
\end{equation}

\begin{figure}
    \centering
    \includegraphics[]{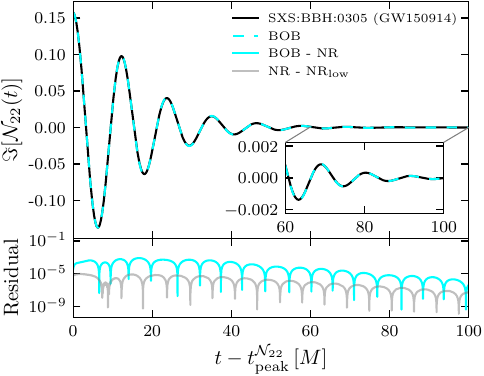}
    \caption{Comparison of BOB and a NR waveform for the imaginary part of the $(2,2)$ mode of the news for an SXS simulation \cite{SXS:BBH:0305,sxs_cat1,SXSCatalogData_3.0.0,sxs_cat3,Boyle2025SXS,0305_lovelace2016modeling} with parameters similar to GW150914 \cite{gw150914}. Eqs.~(\ref{eq:BOB_amplitude}) and (\ref{eq:News_phase}) were used to construct BOB, with $\Omega_0$ determined from the fit in Eq.~\eqref{eq:Omega_0_fit}. The remnant mass and spin from the NR simulation is used to determine the fundamental quasinormal mode used in BOB. In the bottom panel, we plot the residual between BOB and NR as well as the residual between the highest (Lev6) and second highest resolution (Lev5) NR waveforms.}
    \label{fig:BOB_2325}
\end{figure}

In the case of $t_0 = -\infty$, the phase simplifies to
\begin{equation}
\begin{split}
    \Phi &= \frac{\tau}{2}\bigg[\Omega_{\text{QNM}}\ln\bigg(\frac{\Omega + \Omega_{\text{QNM}}}{|\Omega - \Omega_{\text{QNM}}|}\bigg) - \Omega_0 \ln\bigg(\frac{\Omega + \Omega_{\text{0}}}{\Omega - \Omega_{\text{0}}}\bigg)\bigg] + \Phi_0.
\end{split}
\label{eq:News_phase}
\end{equation}
As shown in Fig.~\ref{fig:BOB_2325}, assuming Eq.~\eqref{eq:BOB_amplitude} describes $|$\news$|$ can provide a highly accurate model for the merger-ringdown radiation when compared to NR simulations. 

We emphasize that we now have three different expressions for $\Omega$, and therefore for $\omega$ and $\phi$; rather than introduce subscripts, we will rely on context to make it clear which expressions are being used.\footnote{The choice of G, K, B, F to name the subexpressions is arbitrary and no meaning is to be inferred from the names.} In particular, in Sec.~\ref{sec:results} we will primarily utilize Eqs.~\eqref{eq:BOB_amplitude} and \eqref{eq:News_phase}. Eq.~\eqref{eq:News_phase_finite_t0} is included for completeness and not utilized in this work. Throughout out this work, we relate $\Omega$ and $\Phi$ to the waveform quantities $\omega$ and $\phi$ through the relations $\omega = m\Omega$ and $\phi = m\Phi$, where $m$ = 2 in this paper due to our focus on the $(2,2)$ mode.

\subsection{Obtaining strain from BOB}
\label{sec:strain_from_BOB}
In the cases that we do not assume that the BOB amplitude, Eq.~\eqref{eq:BOB_amplitude}, directly models $|h|$, but rather models $|\mathcal{N}|$ or $|$\psi4$|$, we then need a method to obtain $h$. Because BOB is not a full IMR model, but rather is only accurate from the late inspiral onwards (specifically far from the innermost stable circular orbit, or ISCO), we cannot use BOB to describe the entire inspiral contribution to \strain. To obtain \strain\, from \news, we need to perform a single time integral. Since we have an analytical model for the amplitude and phase of \news, the most accurate approach would be to analytically solve the integral 
\begin{equation}
    h=\int{\mathcal{N}} dt = \int{A e^{i\phi}} dt
    \label{eq:strain_from_news}
\end{equation}
However, due to the complicated nature of $\phi$, we are unable to find a closed form solution. We can find an approximate solution by constructing an asymptotic expansion. Integrating Eq.~\eqref{eq:strain_from_news} by parts yields
\begin{equation}
    \int{Ae^{i\phi}} dt = \frac{A}{i\omega}e^{i\phi} - \int{\frac{d}{dt}\bigg[\frac{A}{i\omega}\bigg]e^{i\phi}dt}
    \label{eq:integrate_by_parts_step1}
\end{equation}
Integrating the remaining integral by parts again yields
\begin{align}
  \frac{A}{i\omega}e^{i\phi} &- \frac{1}{i\omega}\frac{d}{dt} \bigg[\frac{A}{i\omega}\bigg]e^{i\phi} \\&+ \int{\frac{d}{dt}\bigg[\frac{1}{i\omega}\frac{d}{dt} \bigg(\frac{A}{i\omega}\bigg)\bigg]e^{i\phi}dt} \nonumber
   \label{eq:integrate_by_parts_step2}
\end{align}
Continuing the integration by parts will provide an approximate solution to \strain\, in terms of an asymptotic series
\begin{equation}
    h = e^{i\phi}\sum_{n=0}^{\infty} (-1)^n \mathcal{D}^n\bigg(\frac{A}{i\omega}\bigg),
    \label{eq:strain_from_news_series}
\end{equation}
where we define the operator $\mathcal{D}$ as 
\begin{equation}
    \mathcal{D} \equiv \frac{1}{i\omega} \frac{d}{dt}
\end{equation}
We note that Eq.~(\ref{eq:strain_news}) is simply the first term in this series.
For a finite $N$, the full solution can be written in terms of the operator $\mathcal{D}$ as
\begin{align}
    h = &e^{i\phi}\sum_{n=0}^{N} (-1)^n \mathcal{D}^n\bigg(\frac{A}{i\omega}\bigg) +\\ &(-1)^{N+1}\int\frac{d}{dt}\bigg[\mathcal{D}^N\bigg(\frac{A}{i\omega}\bigg)\bigg]e^{i\phi}dt
\end{align}

If we instead assume that the BOB amplitude Eq.~\eqref{eq:BOB_amplitude} models $|$\psi4$|$, we need to integrate twice to obtain \strain. Repeating the procedure from Eqs.~\eqref{eq:integrate_by_parts_step1}-\eqref{eq:strain_from_news_series} a second time, we obtain 
\begin{equation}
    h = e^{i\phi} \sum_{m=0}^{\infty} (-1)^m \mathcal{D}^m \bigg[\frac{1}{i\omega}\sum_{n=0}^{\infty}(-1)^n\mathcal{D}^n\bigg(\frac{A}{i\omega}\bigg)\bigg]
\end{equation}
In Eq.~(\ref{eq:strain_news}), we derive an analytical frequency model by relating $h$ and $\mathcal{N}$ using only the leading order term in this series approximation. Although including additional terms would likely improve the model’s accuracy, retaining only the first term allows for a simpler analytical expression for the frequency.

\subsection{Initial Conditions}
\label{sec:IC}
Accurately implementing BOB depends critically on handling the initial conditions $t_0$ and $\Omega_0$ correctly. The ideal approach to setting these parameters can vary significantly depending on the application and can affect the model's overall accuracy substantially. For instance, when attaching BOB to an inspiral waveform, as is done in \cite{Mahesh:2025}, the attachment procedure may naturally constrain $t_0$ and $\Omega_0$ based on continuity. 

For the inverse problem, recovering source parameters from a known waveform, we have more flexibility. One approach, taken in \cite{universal_superkick}, is to treat BOB as a fully phenomenological model, where even the peak amplitude $A_p$ and peak time $t_p$ are obtained from a least-squares fit to NR data. In this work, we adopt a more physical approach, taking $A_p$ and $t_p$ not as free parameters but as well-defined physical quantities of the waveform itself. Therefore, we do not optimize over these values; we set them directly from the peak of the input waveform. This leaves $t_0$ and $\Omega_0$ as the remaining model choices. While physically motivated choices like $t_0=-\infty$ and $\Omega_0 = \Omega_\mathrm{ISCO}$ or $\Omega_0 = 0$ were possibilities suggested in \cite{BOB}, there is significant freedom in the specification of $t_0$ and $\Omega_0$, since they specify the behavior of the model well before merger, where the model is known to be invalid.

We focus our analysis on ``flavors'' of BOB that take the limit $t_0=-\infty$. Note that $t_0$ appears in the phase evolution through $\tanh(t_0-t_p)/\tau$ so $\Omega$ rapidly approaches its value at $t=-\infty$ going backwards from the merger. We find this simplification is well justified, since treating $t_0$ as an additional free parameter does not meaningfully increase accuracy but does increase the model's complexity. With this choice, $\Omega_0$ becomes the only free initial condition parameter. Given that BOB is inaccurate well before the peak, the correct physical interpretation of $\Omega_0$, the orbital frequency at $t=-\infty$, is unclear. Sensible choices could include zero, the ISCO frequency, or an orbital frequency where radiation reaction becomes negligible.

\begin{figure}
    \centering
    \includegraphics[]{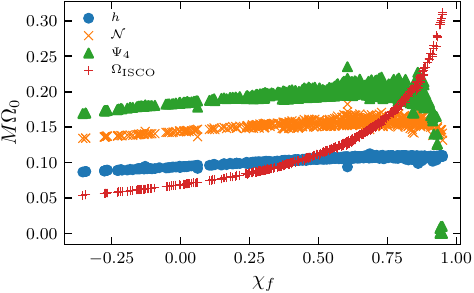}
    \caption{$\Omega_0$ values obtained from a least squares fit of NR frequency data to the respective BOB frequency evolution. For reference, the prograde ISCO frequency of the remnant black hole \cite{bardeen} is provided.} 
    \label{fig:omega0_fit}
\end{figure}

To resolve this ambiguity empirically, we investigate the optimal value of $\Omega_0$ by performing a least-squares fit. For each gravitational wave quantity (e.g., $\Psi_4$), we fit the BOB frequency evolution to the corresponding NR frequency evolution, with $\Omega_0$ as the only free parameter. Fig.~\ref{fig:omega0_fit} shows the best fit $\Omega_0$ values as a function of the remnant black hole's final spin. Surprisingly, we find a near universal value for $\Omega_0$
 across the entire quasi-circular, non-precessing parameter space. To ensure this result is robust against numerical noise at late times, we performed the fit over two intervals, [$t^{{\text{GW}}_{22}}_p$, $t^{{\text{GW}}_{22}}_p + 75M$] and [$t^{{\text{GW}}_{22}}_p$, $t^{{\text{GW}}_{22}}_p + 25M$], where GW represents the appropriate gravitational wave quantity, and found no significant changes. We note some non-linear behavior for \psi4 for very high remnant spins. While a single universal value for $\Omega_0$ is a reasonable approximation, we capture this behavior more precisely with a simple fit of the form:
\begin{equation}
    M\Omega_0 = AM_f + B\chi_f + C,
    \label{eq:Omega_0_fit}
\end{equation}
where $M_f$ is the mass of the remnant black hole in units of $M$, $-1<\chi_f<1$ is its dimensionless spin, and negative values for $\chi_f$ indicate that the remnant spin points in the opposite direction from the initial orbital angular momentum of the binary.
\begin{table}[htb]
\begin{ruledtabular}
\begin{tabular}{lccc}
Quantity & \(A\) & \(B\) & \(C\) \\ \hline
$h$ & 0.01663 & 0.01798 & 0.07883 \\
$\mathcal{N}$   & 0.33568 & 0.03451 & $-0.18763$ \\
$\Psi_4$ & 1.4297 & 0.08424 & $-1.2285$ \\
\end{tabular}
\end{ruledtabular}
\caption{\label{tab:fit_coeffs}Coefficients \(A\), \(B\), and \(C\) in Eq.~\eqref{eq:Omega_0_fit} obtained from fits of NR frequencies to the corresponding BOB frequency evolution.}
\end{table}

\begin{figure*}
    \centering
    \includegraphics[]{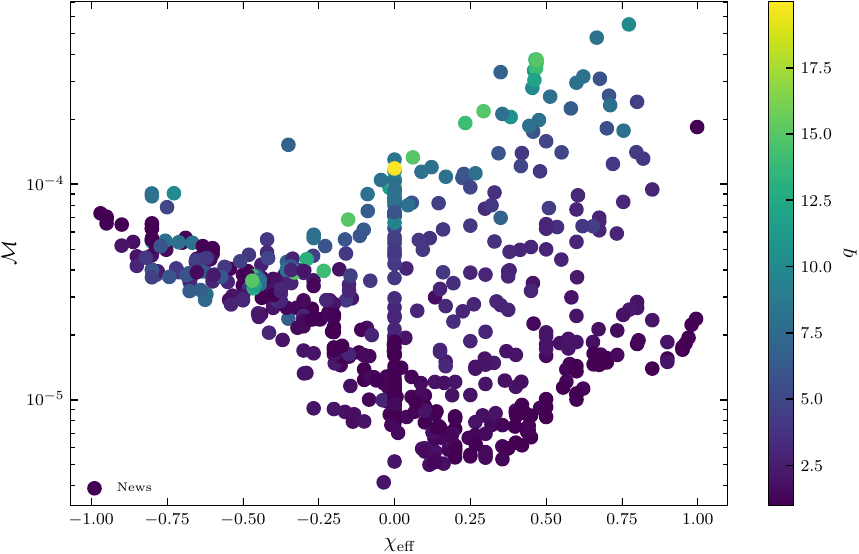}
    \caption{\news\, mismatch between BOB and NR for the $(2,2)$ mode for all quasi-circular and non-precessing systems in the SXS catalog as a function of the initial parameters of the binary. $\Omega_0$ is obtained by performing a least-squares fit to the NR frequency.}
    \label{fig:BOB_news_mismatch_init}
\end{figure*}
\begin{figure*}
    \centering
    \includegraphics[]{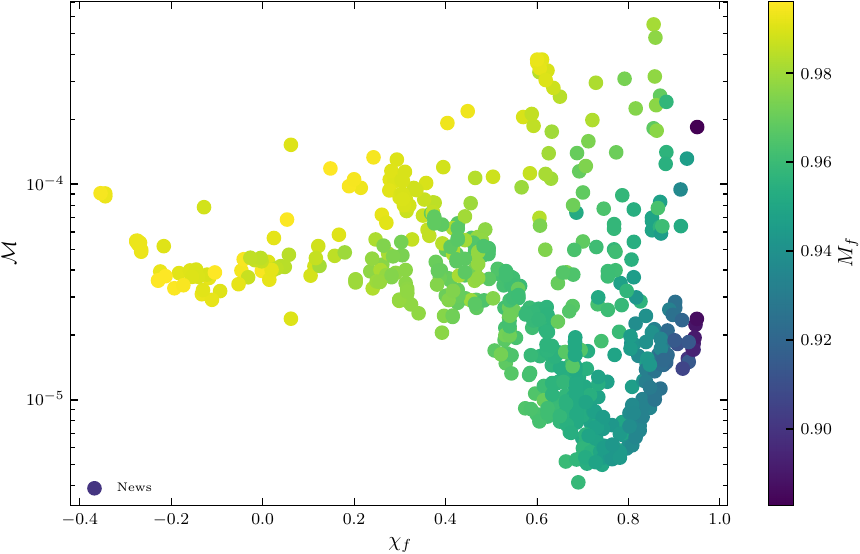}
    \caption{\news\, mismatch between BOB and NR for the $(2,2)$ mode for all quasi-circular and non-precessing systems in the SXS catalog as a function of the remnant parameters of the system. $\Omega_0$ is obtained by performing a least-squares fit to the NR frequency.}
    \label{fig:BOB_news_mismatch_rem}
\end{figure*}
\subsection{What Quantity Does BOB Best Model?}
\begin{figure}
    \centering
    \includegraphics[]{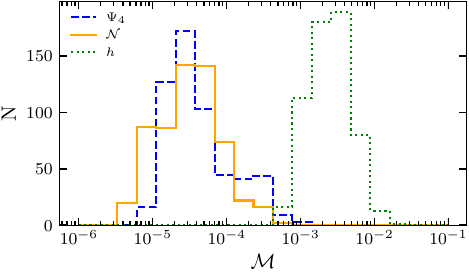}
    \caption{BOB mismatch for the $(2,2)$ mode against the corresponding NR gravitational wave quantity. While BOB can successfully model both \news\, and $\Psi_4$ directly, it does not accurately model \strain\, directly, indicating that the most accurate way to obtain \strain\, is to integrate BOB.}
    \label{fig:BOB3}
\end{figure}
Since the frequency evolution of BOB depends on which quantity (\strain, \news, or \psi4) has its amplitude modeled by Eq.~\eqref{eq:BOB_amplitude}, a critical first step is to determine which quantity, the strain \strain, the news \news, or the Weyl scalar \psi4, BOB most accurately describes. In the original work \cite{BOB}, one of us (STM) suggested that BOB could most accurately describes $\Psi_4$, based on the behavior of perturbations in the limit of a slowly spinning remnant, while the authors of \cite{universal_superkick} used superkick configurations to suggest that BOB most accurately describes $\mathcal{N}$. We approach this question through several viewpoints, and will provide points of evidence that BOB best models \news throughout this work. First, we can directly compare the mismatches obtained by comparison to the SXS catalog when we assume BOB best models $\Psi_4$, \news\, or \strain. We define the mismatch for \strain\, as 
\begin{equation}
    \mathcal{M} = 1-\max_{t,\phi_0}\frac{\braket{h^\mathrm{NR}_{22},h^\mathrm{Model}_{22}}}{\sqrt{\braket{h^\mathrm{NR}_{22},h^\mathrm{NR}_{22}}\braket{h^\mathrm{Model}_{22},h^\mathrm{Model}_{22}}}},
    \label{eq:mismatch}
\end{equation}
where 
\begin{equation}
    \braket{h^\mathrm{NR}_{22},h^\mathrm{Model}_{22}} = \int^{t_f}_{t_0}{\bar{h}^\mathrm{NR}_{22}h^\mathrm{Model}_{22}}\,dt,
\end{equation}
and $\bar{h}$ denotes the complex conjugate of $h$. The mismatch for \news and $\Psi_4$ are defined similarly. We minimize the mismatch over a time and phase shift between the two waveforms. Figs.~\ref{fig:BOB_news_mismatch_init} and \ref{fig:BOB_news_mismatch_rem} show the $\mathcal{N}_{22}$ mismatch of BOB against SXS waveforms as a function of the initial and final parameters of the system, respectively. As listed in Table \ref{tab:fit_mismatch}, allowing $\Omega_0$ to be best fit to the NR frequency, as done in Figs.~\ref{fig:BOB_news_mismatch_init} and \ref{fig:BOB_news_mismatch_rem}, results in a mean \news\, mismatch of $4.61 \times 10^{-5}$ and a median mismatch of $2.85 \times 10^{-5}$ across the SXS catalog for quasicircular and non-precessing cases evaluated between $t_p^\mathcal{N}$ and $t_p^\mathcal{N} + 75M$. Using the fit of Eq.~(\ref{eq:Omega_0_fit}) results in a mean mismatch of $5.17 \times 10^{-5}$ and a median mismatch of $3.18 \times 10^{-5}$.
\begin{table}[h]
\centering
\begin{tabular}{lcc}
\hline
 & Least-squares $\Omega_0$ & Universal $\Omega_0$ fit \\
\hline
Mean          & 4.61 & 5.17 \\
Median        & 2.85 & 3.18 \\
Best (sim)    & 0.414 (SXS:BBH:2126) & 0.365 (SXS:BBH:0448) \\
Worst (sim)   & 55.1 (SXS:BBH:2515)  & 66.6 (SXS:BBH:0204) \\
\hline
\end{tabular}
\caption{
$\mathcal{N}_{22}$ mismatch (in units of $10^{-5}$) when performing a least squares fit of the NR frequency to the BOB frequency, Eq.~(\ref{eq:BOB_news_frequency_evolution}), compared to using the fit in Eq.~(\ref{eq:Omega_0_fit}).}
\label{tab:fit_mismatch}
\end{table}

Fig.~\ref{fig:BOB3} shows that while BOB can model both $\Psi_4$ and \news\, to high accuracy, modeling the amplitude of \strain\, with the fundamental BOB amplitude (Eq.~\ref{eq:BOB_amplitude}) provides a poor approximation. Rather we need to integrate \news\, to obtain \strain. Therefore, to obtain \strain, BOB needs to be integrated. As described in Sec.~\ref{sec:strain_from_BOB}, we are unable to find an exact analytical solution to the integration of BOB, so we resort to an approximate analytical solution represented as an asymptotic series. Because any approximate solution to an integration, be it numerical or analytical, will introduce error, obtaining \strain\, from \news\, provides a practical advantage as it only requires one integration while $\Psi_4$ requires two. In addition to this approximate analytical solution, we tested both time-domain numerical integration, with high and low pass filters, and fixed-frequency integration \cite{FFI}. We found this approach can match or outperform the accuracy of our numerical integration approaches, while retaining the analytical nature of BOB, which is a key advantage of this approach to merger-ringdown modeling.

\section{Results}
\label{sec:results}
 Having determined that BOB most accurately describes $\mathcal{N}$, we compare BOB to state of the art waveform models and probe how BOB can provide insights into the underlying physics of black hole binary mergers. In this work, all our results focus only on the $(2,2)$ mode for quasi-circular and non-precessing systems. Unless specified, our QNM indices will refer to the prograde mode, i.~e.~the perturbation rotating with the spin of the remnant. In cases where both the prograde and retrograde modes are analyzed, the prograde mode is denoted with a ``$+$'' and the retrograde mode is denoted with a ``$-$''. While our overtone analysis will focus on \news, for easier comparison to other studies, many of our results will have zero on the time axis refer to $t^h_p$, the time of the peak $L^2$ norm of \strain.
 
\subsection{Comparing BOB to a sum of overtones}
 A natural benchmark for BOB is a merger-ringdown model built from a sum of quasinormal mode (QNM) overtones \cite{importance_of_overtones,qnmfits,multimode_qnm,improved_qnm_extraction}, as both models depend directly on the remnant black hole's properties. This comparison is further motivated by BOB's physical foundation; its amplitude evolution derives from the perturbation of null geodesics from the remnant's light ring, a concept closely linked to the light-ring correspondence for QNMs \cite{qnm_lightring1,qnm_lightring2,bertiqnmreview}. Although BOB cannot be written directly as a sum of overtones, it shares many similarities that are worth exploring. In particular the amplitude evolution in BOB, given in Eq.~\eqref{eq:BOB_amplitude}, can be rewritten as an infinite series,
 \begin{align}
    A_p\text{sech}\left(\frac{t-t_p}{\tau}\right) = 2 A_p \sum_{n=0}^{\infty} (-1)^n e^{-(2n+1)\frac{t-t_p}{\tau}} \notag \\
    = 2A_p\bigg(e^{-\frac{t-t_p}{\tau}} - e^{-3\frac{t-t_p}{\tau}} + e^{-5\frac{t-t_p}{\tau}} + ...\bigg).
    \label{eq:amplitude_expansion}
\end{align}
This series reveals how BOB's amplitude structure incorporates the overtone damping times, $\tau_n$, in the eikonal limit \cite{bertiqnmreview} where
\begin{equation}
    \tau_n = \frac{\tau_{n=0}}{2n+1}.
\end{equation}

\begin{figure}
    \centering
    \includegraphics[]{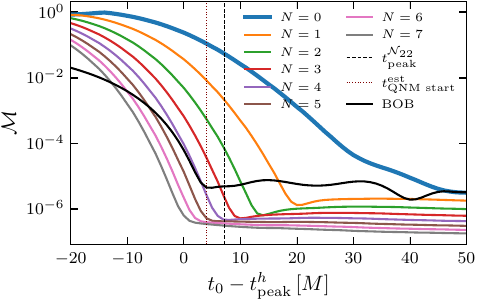}
    \caption{Accumulated mismatch comparisons of $\mathcal{N}_{22}$ between BOB or a sum of QNM overtones and CCE waveforms for an equal mass, $\vec{\chi}_i\cdot\hat{z}=0.2$ aligned spin case, where $\vec{\chi}_i$ is the initial dimensionless spin of each black hole (q1\_aligned\_chi0\_2 in the CCE catalog) and the $r = 261M$ worldtube data is used. For the $\Omega_0$ parameter in BOB, we use the fit given in Eq.~\eqref{eq:Omega_0_fit}. The mismatch is computed from $t_0$ to $t_0 + 100M$. While we compute the mismatch for \news, we provide the time with reference to the peak of the $L^2$ norm of \strain\, for easier comparison with other studies. $t^{\text{est}}_{\text{QNM start}}$ is the earliest time at which stable QNMs could be extracted in \cite{nonlinear_qnm}.}
    \label{fig:BOB_CCE_single}
\end{figure}

To assess the accuracy of BOB compared to a sum of QNMs, in Fig.~\ref{fig:BOB_CCE_single} we plot the accumulated mismatch between BOB and NR for $\mathcal{N}_{22}$ as a function of the initial time $t_0$ used to calculate the mismatch. We calculate the mismatch from $t_0$ until $t_0+100M$. We use the q1\_aligned\_chi0\_2 data available in the public EXT-CCE waveform database \cite{sxs_cat1,sxs_cat2,sxs_cat3,cce_cat,cce_citation,cce_website}, with a worldtube radius of $r = 261M$ transformed to the superrest frame \cite{mitman2021fixing,mitman2022fixing,mitman2024review}. Using the \texttt{qnmfits} \cite{qnmfits} package, we simultaneously fit a large number of overtones using standard non-linear least squares fitting. The QNM and BOB mismatch behaviors are qualitatively similar, each showing a clear ``knee'' after which the mismatch stabilizes. Crucially, BOB maintains higher accuracy than the QNM model prior to the merger peak. For this system, BOB's accuracy at the peak of $\mathcal{N}_{22}$ is comparable to a four-QNM model, and at the peak of the $L^2$ norm of \strain, $t^h_\mathrm{peak}$, BOB's accuracy is comparable to a five QNM model. Therefore, matching BOB's accuracy using a sum of QNMs requires fitting 8-10 coefficients directly to the NR data. Because we use the fit for $\Omega_0$ described in Eq.~\eqref{eq:Omega_0_fit}, BOB has no other parameters that need to be fit because in both the QNM and BOB case the final mass and spin are set from the NR simulation values.

\begin{figure}
    \centering
    \includegraphics[]{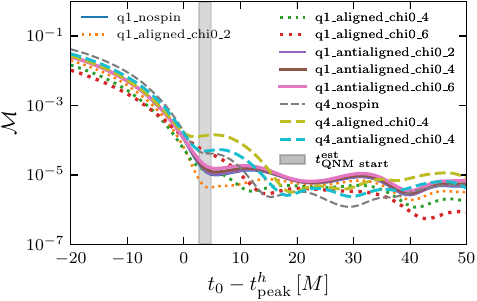}
    \caption{Accumulated mismatch comparisons of $\mathcal{N}_{22}$ between BOB and CCE waveforms for all non-precessing cases in the catalog. For the $\Omega_0$ parameter in BOB, we use the fit given in Eq.~\eqref{eq:Omega_0_fit}. $t_\text{QNM start}^\text{est}$ is a rough estimate for the onset of QNM stability in the $(2,2)$ mode for these 10 configurations obtained through the \texttt{qnmfinder} code with parameters similar to \cite{improved_qnm_extraction}. $q=4$ systems are shown with dashed lines, equal mass systems with $\chi_\mathrm{eff} = 0$ are shown with solid lines, and equal mass systems with $\chi_\mathrm{eff} > 0$ are shown with dotted lines. `Aligned' refers to systems where the initial spins of both black holes are aligned with the initial orbital angular momentum while `Antialigned' refers to systems where the initial spin of the secondary black hole is pointed opposite to the initial orbital angular momentum.} 
    \label{fig:BOB_CCE_all}
\end{figure}

The ``knee'' in the curves in Figure~\ref{fig:BOB_CCE_single} where BOB's mismatch stabilizes does not appear to be arbitrary; it correlates to within $1M$ with the earliest time that stable QNMs could be extracted for this system in \cite{nonlinear_qnm}, suggesting this knee for BOB's mismatches may correspond to the physically meaningful ringdown start time. Fig.~\ref{fig:BOB_CCE_all} extends this analysis to all non-precessing CCE configurations, which are transformed to the superrest frame, and are available in the public EXT-CCE catalog. Using the \texttt{qnmfinder} code we obtain a rough estimate for the onset of QNM stability in the $(2,2)$ mode for these ten configurations, which again generally coincides with the knee seen in BOB mismatches, further suggesting that BOB may be useful in determining the start of the linear ringdown. We also note that even before the ``knee'', BOB's mismatches at earlier times grow more slowly than a sum of the dominant mode and any number of overtones, due to BOB's ability to capture the ``direct wave'' portion of the merger radiation \cite{kankani2026direct,oshita2025probing}, further suggesting that BOB provides a superior description of the waveform.

Perturbation theory tells us that the amplitude of each overtone is directly proportional to a source dependent term, which contains a dependence on the overtone index $n$ (i.e.  how the ``bell'' is struck affects each overtone differently) \cite{berti2006quasinormal,bertiqnmreview,zhang2013quasinormal}. A standard QNM model must fit for the amplitude of each overtone independently. In contrast, BOB's construction requires no such fitting. As seen in Eqs.~\eqref{eq:BOB_amplitude} and \eqref{eq:amplitude_expansion}, while BOB incorporates information from an infinite number of eikonal overtones in its amplitude evolution, the coefficients attached to each overtone-like term in the summation differ only by an overall alternating sign and are all scaled by the single peak amplitude $A_p$. This implies that for systems with identical remnant properties, the merger-ringdown radiation will differ only, to a very accurate approximation, by an overall amplitude scaling. BOB predicts that the perturbation to the remnant black hole, specifically the spatial integral over the source term $I(\omega, \,r)$ \cite{berti2006quasinormal,bertiqnmreview}, is largely independent of the overtone index \textit{n} and is encoded in the peak amplitude of the waveform. This prediction is consistent with observations made in several studies that have fit QNMs to NR waveforms for non-precessing systems \cite{nonlinear_qnm,improved_qnm_extraction,importance_of_overtones,cheung2024extracting}. Furthermore, other numerical studies have independently shown that remnant properties appear to be strongly correlated with quantities measured at the waveform's peak \cite{amp_rem1,amp_rem2}, further suggesting that remnant information is encoded in the peak amplitude. We stress that while these observations have all required the use of numerical relativity simulations, this prediction emerges naturally from the BOB approach to merger-ringdown modeling. 

Lastly, it is important to note that BOB does not reproduce the full Kerr excitation structure. The approximate n-independence of $I(\omega_n,\, r)$ is what makes an analytic resummation of the net QNM contribution possible, and that the success of this resummation in matching NR is itself evidence supporting that approximation. BOB will not give the exact excitation factors for each overtone. It will reproduce the combined effect of the overtones. Therefore, BOB is not a discrete sum of QNMs but a resummation of an infinite tower of them, and these two objects are not directly comparable coefficient-by-coefficient until late times when only the fundamental QNM remains.

\subsubsection{Analytically Recovering $\mathcal{A}_{220}$}
\begin{figure}
    \centering
    \includegraphics[width=0.45\textwidth]{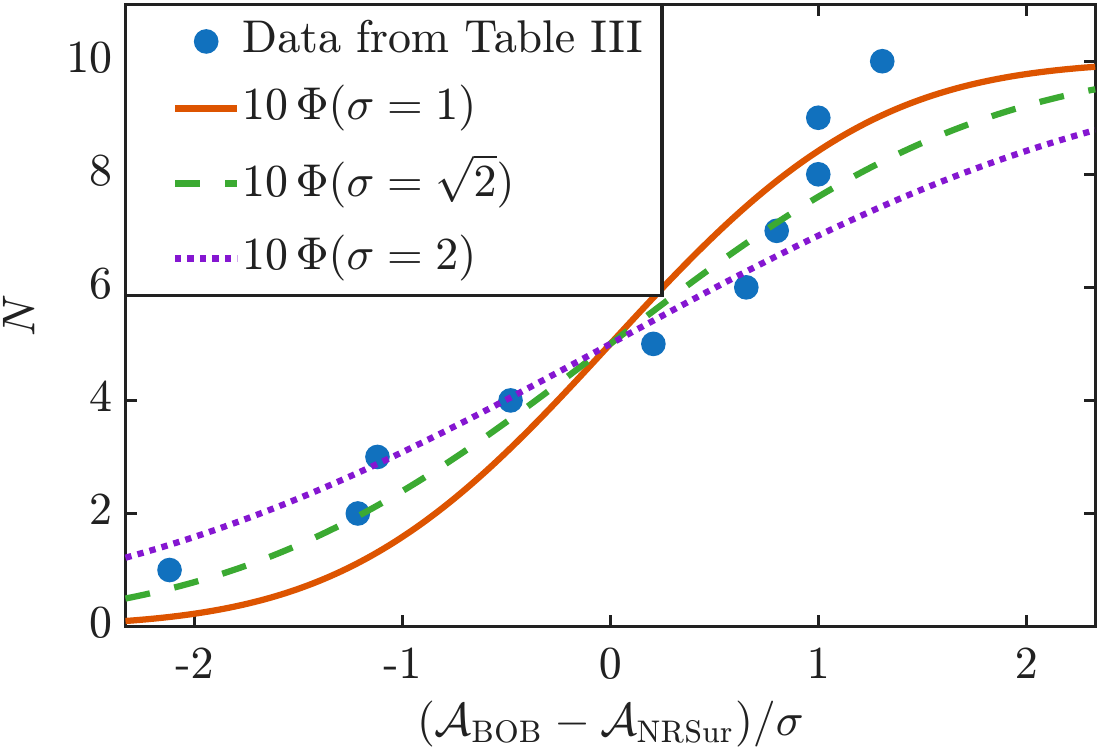}
    \caption{Cumulative number of points, $N$, from Table \ref{tab:CCE_A220_diff} (blue dots) with standardized errors as large as $|\mathcal{A}_{\text{BOB}}-\mathcal{A}_{\text{NRSur}}|/\sigma$  compared to expected cumulative values for a normal distribution with $\sigma=1$ (as would be expected if the only source of error is $\mathcal{A}_{\text{NRSur}}$, solid orange), $\sigma=\sqrt{2}$ (as would be expected if the error in $\mathcal{A}_{\text{BOB}}$ followed the same distribution as $\mathcal{A}_{\text{NRSur}}$, dashed green), and $\sigma=2$ (as would be expected if the error in $\mathcal{A}_{\text{BOB}}$ followed a distribution whose standard deviation is $\sqrt{3}$-times as large as $\mathcal{A}_{\text{NRSur}}$, dotted purple). The data is consistent with errors in $\mathcal{A}_{\text{BOB}}$ being at worst comparable in size to the errors in $\mathcal{A}_{\text{NRSur}}$.}
    \label{fig:A220err}
\end{figure}

A key prediction of BOB, first noted in \cite{universal_superkick}, is the analytic link between the peak amplitude of the waveform and the amplitude of the fundamental QNM, $\mathcal{A}_{220}$, when describing the strain.\footnote{Throughout this section, we extract QNM amplitudes from the news, but we report the corresponding strain amplitude, obtained through $\mathcal{A}_{\text{strain},22n} = \mathcal{A}_{\text{news},22n}/|\omega_{22n}|$ for easier comparison to other works.} This connection arises from matching the late time behavior of BOB to that of the fundamental mode. The precise form of this relationship depends on which gravitational wave quantity BOB models. In \cite{universal_superkick}, the `SKD4-03' superkick configuration was studied and it was found that BOB's peak amplitude was best able to predict the amplitude of the fundamental mode when the BOB amplitude function, Eq.~\eqref{eq:BOB_amplitude}, was used to model the amplitude of $|$\news$|$, resulting in the following relationship:
\begin{equation}
    \mathcal{A}_{220} (t) = \frac{2A_{\text{peak}_{\mathcal{N}_{22}}}}{|w_{220}|} e^{\frac{-(t-t_p)}{\tau}},
    \label{eq:BOB_A220}
\end{equation}
where $w_{220}$ is the complex quasinormal mode frequency. While BOB's amplitude evolution incorporates overtones through the eikonal limit, we use the true Kerr QNM, $\omega_{220}$, for the fundamental mode. This allows for a direct match to a single QNM at late times. In this work we evaluate the QNM amplitudes at $t_p^h + 20M$. We test the accuracy of this analytic prediction across the full non-precessing and quasi-circular SXS catalog and confirm that \news\, best satisfies the relationship between the peak amplitude and the amplitude of the fundamental quasinormal mode. Fig.~\ref{fig:A2201_sxs} compares the BOB predicted value for $\mathcal{A}_{220}$ to the value obtained by fitting a QNM model, consisting of two progade and one retrograde mode, directly to the NR simulations. For simplicity, we only show modes for systems where the final spin is either zero or in the same direction as the binary’s initial orbital angular momentum, i.e., $\vec{\chi}_f \cdot \hat{z} \geq 0$ for these simulations. Because a large CCE waveform database is not currently public, we use the non-CCE SXS waveform database to study the parameter space comprehensively. Therefore, the NR value for $\mathcal{A}_{220}$ may contain a small bias. For a more precise comparison using superrest frame waveforms, we turn to the ten publicly available non-precessing CCE configurations. In Table~\ref{tab:CCE_A220_diff}, we compare the BOB prediction of $\mathcal{A}_{220}$ to a value obtained from a specialized ringdown surrogate model \cite{surr_ringdown}. The results are remarkably similar. In fact, if we take the difference between the two predictions, standardize it (i.~e.~normalize it by the $1\sigma$ errors of the surrogate), and plot the cumulative number of points with standardized errors at least as large as $|\mathcal{A}_{\text{BOB}}-\mathcal{A}_{\text{NRSur}}|/\sigma$, the result is consistent with the underlying errors in BOB being, at worst, comparable in size to, and possibly much smaller than, the errors in NRSur3dq8\_RD \cite{surr_ringdown}, see Figure \ref{fig:A220err}. 

Requiring only the peak $\mathcal{N}_{22}$ amplitude and the remnant properties ($M_f,\chi_f$), BOB's prediction of the fundamental mode amplitude is, at worst, comparably accurate to ringdown surrogate models. This connection between the peak amplitude and the fundamental mode further emphasizes the argument that the perturbation encoded during the non-linear merger is largely independent of the overtone index \textit{n}. The ability to analytically link a strong field result to the late ringdown further emphasizes how the BOB approach to merger-ringdown modeling can help better illuminate the underlying physics of black hole mergers.

\begin{table}[h!]
\centering
\begin{ruledtabular}
\begin{tabular}{cccc}
ID & $\mathcal{A}_{\text{BOB}}$ & $\mathcal{A}_{\text{NRSur}}\pm \sigma$ & $|\mathcal{A}_{\text{BOB}}-\mathcal{A}_{\text{NRSur}}|/\sigma$ \\
\hline
1  & 0.1811 & $0.1785 \pm 0.0026$ & 1.0000 \\
2  & 0.1943 & $0.1926 \pm 0.0026$ & 0.6538 \\
3  & 0.2114 & $0.2126 \pm 0.0025$ & 0.4800 \\
4  & 0.2378 & $0.2431 \pm 0.0025$ & 2.1200 \\
5  & 0.1817 & $0.1783 \pm 0.0026$ & 1.3077 \\
6  & 0.1802 & $0.1777 \pm 0.0025$ & 1.0000 \\
7  & 0.1788 & $0.1768 \pm 0.0025$ & 0.8000 \\
10 & 0.1032 & $0.1026 \pm 0.0029$ & 0.2069 \\
11 & 0.1210 & $0.1244 \pm 0.0028$ & 1.2143 \\
12 & 0.1183 & $0.1211 \pm 0.0025$ & 1.1200 \\
\end{tabular}
\end{ruledtabular}
\caption{Comparison of $\mathcal{A}_{220}$ values obtained from Eq.~\eqref{eq:BOB_A220} with those from a ringdown-specific surrogate model, NRSur3dq8\_RD \cite{surr_ringdown}, for the 10 non-precessing cases publicly available in the EXT-CCE database \cite{cce_website}. Also provided are the $1\sigma$ errors on $\mathcal{A}_{220}$ provided by the surrogate model, and the difference between the BOB and surrogate values divided by those $1\sigma$ surrogate errors. This difference is fully consistent with ten draws from a Gaussian distribution with the stated $1\sigma$ errors on the surrogate, which suggests that BOB is providing the more accurate estimate.}
\label{tab:CCE_A220_diff}
\end{table}

\subsubsection{The Importance of Retrograde Modes}
\begin{figure}
    \centering
    \includegraphics[]{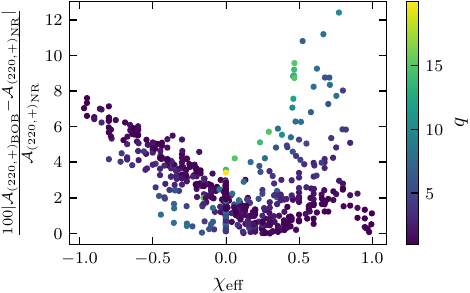}
    \caption{Percent difference between BOB's prediction for the fundamental mode ($l=2$, $m=2$, $n=0$, $+$ refers to prograde), given by Eq.~\eqref{eq:BOB_A220}, and those obtained through fitting to the SXS catalog. }
    \label{fig:A2201_sxs}
\end{figure}

In Fig.~\ref{fig:A2201_sxs}, for systems with $\chi_\text{eff} \lsim 0$, we observe a linear trend in BOB's accuracy as a function of $\chi_\text{eff}$. We observe a similar trend in Fig.~\ref{fig:BOB_news_mismatch_init}. This is intriguing given that BOB depends only on the remnant parameters while $\chi_\text{eff}$ is a measure of the initial state of the binary given by
\begin{equation}
    \chi_\mathrm{eff} = \frac{\chi_1\hat{\boldsymbol{S}}_1 + q\chi_2\hat{\boldsymbol{S}}_2}{1+q} \cdot \hat{\mathbf{L}},
\end{equation}
where $\chi_{1,2}$ are the initial dimensionless spin magnitudes, $\boldsymbol{\hat{S}}_{1,2}$ are the unit spin vectors, $q$ is the initial mass ratio of the binary, and $\mathbf{L}$ is the Newtonian orbital angular momentum vector.
We show that this trend points directly to the dominant missing physics in the BOB model, an insight made possible by the model's minimal tuning to NR.

The final spin of the remnant black hole can be well approximated as the sum of the orbital angular momentum of a particle with mass equal to the reduced mass located at the ISCO of the remnant, and the total spin angular momentum of the system \cite{final_mass}. For $\chi_\mathrm{eff}>0$ these two components add constructively, leading to a highly spinning remnant. For $\chi_\mathrm{eff}<0$, they compete, resulting in a remnant with a lower spin magnitude.

This final spin is critical because it governs the properties of the black hole's QNMs. For any given $(l,m,n)$ QNM index, the Teukolsky equation \cite{teukolsky1973perturbations,bertiqnmreview} predicts two families of modes: prograde (co-rotating) and retrograde (counter-rotating) \cite{cheung2024extracting,berti2006gravitational}. While the prograde modes are typically much more strongly excited in a merger, the relative importance of the retrograde modes increases as the remnant spin decreases. While the BOB formalism should in principle be equally applicable to prograde and retrograde modes, it has only so far been used to describe prograde modes. 

\begin{figure}
    \centering
    \includegraphics[]{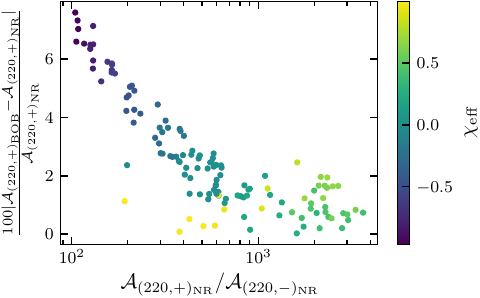}
    \caption{BOB's accuracy at recovering the amplitude of the fundamental prograde mode, $\mathcal{A}_{(220,+)}$ as a function of the relative strength of the prograde and retrograde modes for equal mass systems. The colorbar represents the initial effective spin of the binary, emphasizing that systems with lower $\chi_\text{eff}$ values will excite retrograde modes more strongly.}
    \label{fig:BOB_retrograde_correlation_q1}
\end{figure}

In Fig.~\ref{fig:BOB_retrograde_correlation_q1}, restricted to equal mass binaries to isolate the effect of $\chi_{\text{eff}}$, we plot BOB's accuracy as a function of the relative strength of the prograde mode compared to the retrograde mode. To determine the NR QNM amplitudes, we use \texttt{qnmfits} and fit two prograde and one retrograde mode. For negative values of $\chi_\text{eff}$, the retrograde mode becomes increasingly important. Simultaneously, the accuracy of BOB's prediction of $\mathcal{A}_{(220,+)}$ degrades as $\chi_\text{eff}$ becomes increasingly negative. While the retrograde mode's amplitude remains subdominant, it is important enough to impact the overall accuracy of BOB.  While we leave the incorporation of retrograde modes into BOB for future work, which will greatly benefit from a large public CCE waveform database, this analysis highlights a unique advantage of BOB. Because BOB so accurately captures the prograde physics, the residual between BOB and NR data provides a cleaner signal of the additional retrograde and nonlinear physics; this cleaner signal could potentially allow for the robust extraction of those additional contributions without the risk of overfitting associated with more complex models, and without being obscured by the much louder prograde content. 

\subsubsection{QNM Content in BOB}

BOB exhibits many similarities to quasinormal modes (QNMs). The underlying physics of the BOB model is based on null geodesics perturbed from the remnant light ring. Furthermore, BOB can analytically link the peak amplitude to the fundamental mode amplitude with high accuracy and incorporate information from an infinite number of overtones in its amplitude evolution. BOB also predicts that the source perturbation is largely independent of the overtone index, a finding supported by recent studies \cite{improved_qnm_extraction,cheung2024extracting,nonlinear_qnm,importance_of_overtones}. Additionally, as shown in Fig.~\ref{fig:BOB_CCE_single}, there are qualitative similarities in the mismatch behavior between the two models. Given these striking similarities, the relationship between BOB and QNMs warrants further investigation. While we leave a detailed study for future work, we present some initial findings here.

An obvious question to ask is, what QNM content does BOB contain? We approach this question empirically by fitting QNMs to BOB directly. We take two different approaches for fitting quasinormal modes to BOB, reflecting common practices in the field. First, we perform a standard non-linear least-squares fit using the \texttt{qnmfits} package \cite{qnmfits}. Second, we use the more robust \texttt{qnmfinder} code \cite{improved_qnm_extraction}, which employs the \texttt{varpro} algorithm \cite{varpro1} and performs stability checks to guard against overfitting. As we will show, these two methods yield significantly different results, highlighting a powerful application for BOB: serving as an analytic testbed to assess the robustness of QNM fitting algorithms.

\begin{figure}
    \centering
    \includegraphics[]{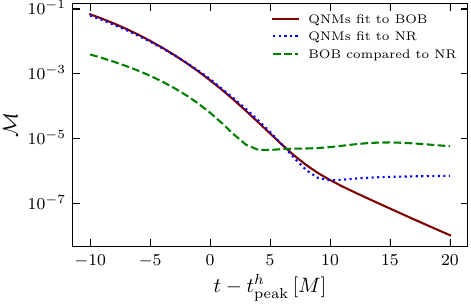}
    \caption{The mismatch for $\mathcal{N}_{22}$ obtained from fitting 4 QNMs to BOB for the superrest transformed q1\_aligned\_chi0\_2 system in the CCE catalog (solid maroon line) compared to the mismatch for 4 QNMs fit directly to the NR data (solid blue line) as well as the mismatch of BOB compared to the NR data (green dashed line).}
    \label{fig:fit_QNM_to_BOB_mismatch}
\end{figure}
\begin{figure}
    \centering
    \includegraphics[]{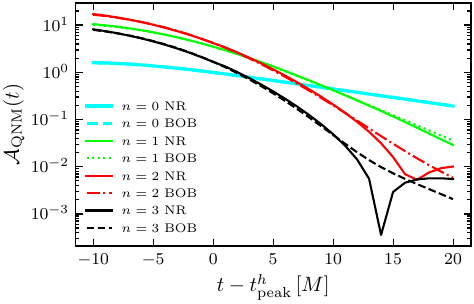}
    \caption{We plot the amplitudes of the QNMs obtained from a standard non-linear least squares fit to \news\, for BOB (dashed) and to the NR news (solid) for the superrest transformed q1\_aligned\_chi0\_2 system in the CCE catalog. The $n=0$ lines lie on top of each other.} 
    \label{fig:fit_QNM_to_BOB_amplitudes}
\end{figure}

We focus our analysis on the q1\_aligned\_chi0\_2 system, whose parameters are almost identical to the system analyzed in \cite{nonlinear_qnm}, and is analyzed in Fig.~\ref{fig:BOB_CCE_single}. We perform a four QNM fit to both the analytic BOB news waveform and the superrest frame NR news. Because BOB does not contain the numerical noise present in all NR waveforms, as seen in Fig.~\ref{fig:fit_QNM_to_BOB_mismatch}, the QNMs are able to continuously obtain lower mismatches as we go further from the peak. As seen in Fig.~\ref{fig:fit_QNM_to_BOB_amplitudes}, the fitted overtone amplitudes from BOB and NR are nearly identical at the time of the peak of the $L^2$ norm of \strain, agreeing to within 1\%. The amplitudes show excellent agreement until a few $M$ after the peak of the waveform, after which the higher overtones quickly diverge. Before and very close to the peak of the waveform, we do not expect quasinormal modes to have yet been excited. Therefore, the agreement before the peak does not imply a true QNM presence, but rather demonstrates that both the BOB and NR waveforms share a very similar morphology during the merger, which the QNM basis is able to capture, leading to nearly identical coefficients. The amplitude divergence $\sim5$--$10M$ after the peak for higher overtones is expected, as they are likely overfitting by this time. However, the strong agreement near the peak further reinforces the idea that BOB contains a significant amount of overtone information, and suggests that BOB can be used to obtain high accuracy estimates for the overtone amplitudes without requiring fitting to NR data; this should prove useful for a variety of fitting routines.

A more robust extraction of QNMs from BOB can be done using the \texttt{qnmfinder} code \cite{improved_qnm_extraction} which takes into account the stability of the modes and utilizes a more robust fitting algorithm. Using similar parameters to those in \cite{improved_qnm_extraction}, we are able to extract the $n=0$ fundamental mode and the $n=1$ overtone from BOB. Extracting the $n=2$ overtone from \texttt{qnmfinder} when analyzing BOB requires loosening the stability criterion by increasing the $\sigma^\text{max}$ parameter to 0.3.

\begin{table}[t]
  \centering
  \begin{ruledtabular}
  \begin{tabular}{llcc}
    Waveform & Method & $\mathcal{A}_{220}$ & $\mathcal{A}_{221}$ \\
    \hline
    BOB & \texttt{qnmfinder}         & 0.1945 & 0.03498 \\
    BOB & \texttt{qnmfits} (2 modes) & 0.1948 & 0.03524 \\
    BOB & \texttt{qnmfits} (4 modes) & 0.1949 & 0.03683 \\
    NR  & \texttt{qnmfinder}         & 0.1933 & 0.03139 \\
    NR  & \texttt{qnmfits} (2 modes) & 0.1933 & 0.02849 \\
    NR  & \texttt{qnmfits} (4 modes) & 0.1936 & 0.03251 \\
  \end{tabular}
  \end{ruledtabular}
\caption{Fundamental mode ($\mathcal{A}_{220}$) and first overtone amplitudes ($\mathcal{A}_{221}$) for the superrest
transformed q1\_aligned\_chi0\_2 configuration, corresponding to ID 2 in Table~\ref{tab:CCE_A220_diff}, extracted at
$t_p^h + 20M$ from the BOB and NR news waveform using \texttt{qnmfinder} and
\texttt{qnmfits}. For $\texttt{qnmfits}$ we fit two cases, one with two prograde modes and one with four prograde modes.}
  \label{tab:qnmfinder_qnmfits_amplitudes}
\end{table}

Table~\ref{tab:qnmfinder_qnmfits_amplitudes} compares the amplitudes of the fundamental mode and the first overtone recovered by \texttt{qnmfinder} and \texttt{qnmfits} from both the BOB and the corresponding NR waveform. For \texttt{qnmfits} we fit two and four prograde QNMs. For BOB, the two codes agree well on the fundamental mode amplitude, and the recovered value is consistent with the analytic prediction of Eq.~\eqref{eq:BOB_A220}. The fundamental mode amplitude recovered from NR is likewise consistent between the codes, and lies within $1\sigma$ of the value obtained for NR by the ringdown-specific surrogate model (Table~\ref{tab:CCE_A220_diff}). The amplitude of the first overtone shows slightly more scatter for both BOB and NR.

If we apply the same process directly to the NR data, we are able to recover the $n=2$ overtone with standard parameters, but are unable to recover higher overtones. For both BOB and NR, stable QNMs first appear around $4M$ after the peak of the $L^2$ norm of \strain, consistent with what was found in \cite{nonlinear_qnm} for the NR data and Fig.~\ref{fig:BOB_CCE_single} for BOB. This alignment further suggests that BOB's structure can help identify the physical onset of the linear ringdown.

While the simple least-squares fit suggests BOB contains a large number of overtones, the more robust algorithm confidently finds only two to three, but is also only able to find three overtones in the NR data. Given BOB's known analytic structure, its high accuracy in the $0$--$10M$ window, and the amplitude agreement seen in Fig.~\ref{fig:fit_QNM_to_BOB_amplitudes}, it is probable that higher overtone content exists in BOB that the qnmfinder algorithm, in its standard configuration, cannot robustly extract. This may indicate that current stability checks are too conservative and could be discarding physical information present in both BOB and the NR data. The analytic and noise-free nature of BOB should make it an ideal laboratory for rigorously testing and improving the techniques used to retrieve higher overtones from gravitational wave signals. For example, one could test whether techniques like mode-dropping, which have been able to extract numerous overtones from this same NR configuration \cite{nonlinear_qnm}, can also robustly extract higher overtone content in BOB, helping to distinguish between physical extraction and overfitting.

\subsubsection{Recovering Remnant Parameters}
\begin{figure}
    \centering
    \includegraphics[]{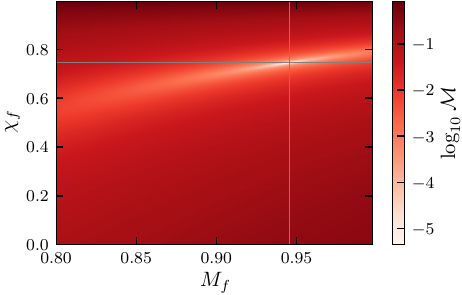}
    \caption{$\mathcal{N}_{22}$ mismatches between BOB and NR (q1\_aligned\_chi0\_2) for a grid of $M_f,\chi_f$ values evaluated in the time interval $[t_p^{\mathcal{N}_{22}},75M+t_p^{\mathcal{N}_{22}}]$. For $\Omega_0$ we use the fit described in Eq.~\eqref{eq:Omega_0_fit}. The intersecting grey lines indicate the NR remnant parameters.}
    \label{fig:CCE2_heatmap}
\end{figure}

\begin{figure}
    \centering
        \includegraphics[]{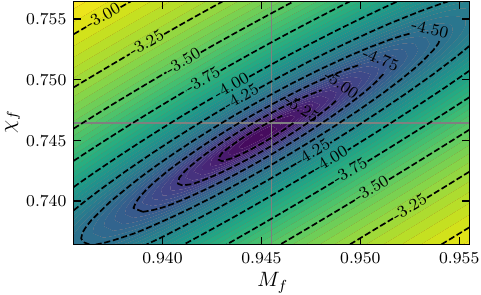}
    \caption{Contour map showing the $\mathcal{N}_{22}$ mismatches between BOB and NR (q1\_aligned\_chi0\_2) for a zoomed in grid of $M_f,\chi_f$ values evaluated in the time interval $[t_p^{\mathcal{N}_{22}},t_p^{\mathcal{N}_{22}} + 75M].$ For $\Omega_0$ in BOB we use the fit described in Eq.~\eqref{eq:Omega_0_fit}. The intersecting grey lines indicate the NR remnant parameters.}
    \label{fig:CCE2_contourmap}
\end{figure}

A key metric for any merger-ringdown model is its ability to accurately recover the remnant mass, $M_f$, and remnant spin, $\chi_f$, from a waveform. We quantify this using a simple metric for the error, $\epsilon$, similar to what is used in \cite{importance_of_overtones,multimode_qnm,surr_ringdown}, defined as: 
\begin{equation}
    \epsilon = \sqrt{(\delta M_f/M)^2 + (\delta \chi_f)^2}.
    \label{eq:simple_erorr}
\end{equation}
Figs.~\ref{fig:CCE2_heatmap} and \ref{fig:CCE2_contourmap} show that the BOB mismatch surface has a well-defined minimum that corresponds closely to the true NR remnant parameters. In Table~\ref{table:CCE_parameter_estimation}, we show the recovered error using BOB against CCE waveforms in the superrest frame over three different time intervals. Additionally, we find a small improvement in the accuracy of the recovered parameters when using BOB to construct \news\, compared to $\Psi_4$, similar to the slightly better mismatch seen in Fig.~\ref{fig:BOB3}, and do not find any clear trend for the parameter recovery accuracy as a function of initial or remnant parameters. To test BOB in a form that only depends on the final mass and spin, we use our fit for $\Omega_0$ (Eq.~\ref{eq:Omega_0_fit}) when the analysis starts at or after the peak of \news. Only for the third column, starting at the peak of the $L^2$ norm of \strain, do we allow $\Omega_0$ to be a free variable. Across the time interval [$t^h_{{p}}$, $t^{\mathcal{N}_{22}}_{p} + 75M$] , BOB achieves a median error of $8.84 \times10^{-3}$. This performance is comparable to that of a 4--8 QNM model \cite{multimode_qnm,importance_of_overtones}, which would require fitting 8--16 free parameters to the waveform, in addition to searching for the true remnant parameters. BOB, on the other hand, has at most one free parameter outside the true remnant parameters.

There is a significant caveat in the use of the simple error as a measure of the accuracy of merger-ringdown models; the accuracy of parameter recovery is highly sensitive to the chosen time interval. As shown in Table.~\ref{table:CCE_parameter_estimation}, the most accurate results are achieved by analyzing only the late-time ringdown, where the waveform has simplified to the fundamental mode. Furthermore, we find that the error $\epsilon$ can vary significantly based on small changes in our fitting methodology. For example, in cases where $\Omega_0$ is a free parameter, our results can change depending on whether we take $\Omega_0$ to be an additional free parameter in an optimization algorithm or if we obtain $\Omega_0$ by performing a least squares fit to the NR waveform. Therefore, $\epsilon$ should be interpreted as a useful benchmark, but not a definitive measure of a model's physical accuracy.

\begin{figure}
    \centering
    \includegraphics[]{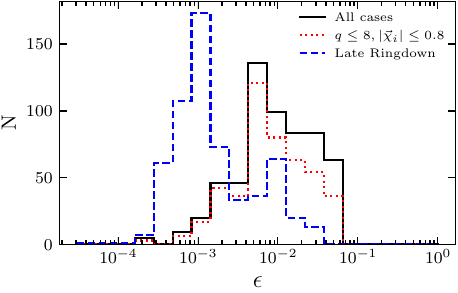}
    \caption{Error in parameter recovery, defined in Eq.~\eqref{eq:simple_erorr} using BOB compared to all quasi-circular and non-precessing cases in the SXS catalog. We plot the error when starting the waveform at the peak of the $L^2$ norm of \strain\, and allow $\Omega_0$ to be fit to the waveform (black). We show this same result, but limited to configurations with $q \leq 8$ and $|\chi_i| \leq 0.8$ (dotted red). Lastly, we show the error obtained when we only use the late ringdown ([$t_p + 40M,t_p + 75M$]) portion of the waveform and use the fit described in Eq.~\eqref{eq:Omega_0_fit} for $\Omega_0$ (dashed blue).} 
    \label{fig:news_parameter_estimation_sxs}
\end{figure}

As shown in Fig.~\ref{fig:news_parameter_estimation_sxs}, when compared to the entire quasi-circular and non-precessing SXS catalog, and fitting $\Omega_0$ to the NR waveform, BOB obtains a median error of $8.77\times10^{-3}$, which again places BOB's performance between the four QNM and eight QNM results presented in \cite{importance_of_overtones,multimode_qnm}. If we limit ourselves to the late ringdown, and use the fit in Eq.~\eqref{eq:Omega_0_fit} for $\Omega_0$, BOB obtains a median error of $1.2 \times 10^{-3}$, which outperforms 8 QNM models that use the entire merger-ringdown. We stress again that the BOB results require, at most, only one parameter fit to the NR waveform, while an overtone model requires two free parameters for each overtone. Furthermore, the results in \cite{importance_of_overtones,multimode_qnm} are limited to cases with $q \leq 8$ and $|\vec{\chi}_i| \leq 0.8$ while as we show in Fig.~\ref{fig:news_parameter_estimation_sxs}, we see the same level of accuracy for cases within and outside these limits with BOB.

\begin{table}[h!]
\centering
\begin{ruledtabular}
\begin{tabular}{c c c c c}
\textbf{Start time:} &$t^{\mathcal{N}_{22}}_{p} + 40M$ & $t^{\mathcal{N}_{22}}_{p, \, \Omega_0 \, \text{fit}}$ & $t^{\mathcal{N}_{22}}_{p, \, \Omega_0 \, \text{free}}$ & $t^h_{{p}}$\\
\hline
CCE ID $=1$  & 1.25  & 7.46  & 12.08 & 0.78 \\
2  & 0.829 & 1.49  & 7.70 & 6.34  \\
3  & 0.528 & 2.17  & 5.09 & 11.32 \\
4  & 0.339 & 3.57  & 3.07 & 11.55 \\
5  & 1.32  & 7.57  & 12.04 & 0.71 \\
6  & 1.43  & 8.33  & 11.27 & 0.67 \\
7  & 1.17  & 9.23  & 11.08 & 0.66 \\
10 & 10.99 & 10.18 & 38.25 & 20.56\\
11 & 4.51  & 19.25 & 9.76 & 31.59\\
12 & 3.59  & 9.48  & 10.91 & 27.95\\ \hline
\textbf{Median:} & \textbf{1.29} & \textbf{8.85} & \textbf{11.00} & \textbf{8.84} \\
\textbf{Mean:} & \textbf{2.6} & \textbf{7.87} & \textbf{12.13} & \textbf{11.22} \\
\end{tabular}
\end{ruledtabular}
\caption{Error $\epsilon$ in recovering the remnant mass and spin, divided by $ 10^{-3}$. This error is defined in Eq.~\eqref{eq:simple_erorr}, and is computed using BOB for the ten non-precessing systems available in the EXT-CCE public database \cite{cce_website}. All CCE waveforms are transformed to the superrest frame. We show the errors obtained from three different time intervals, with the start times indicated at the top, and an end time of $t^{\mathcal{N}_{22}}_{p} + 75M$ for all three cases. For $\Omega_0$ in BOB, we use the fit in Eq.~\eqref{fig:omega0_fit} for the first two columns, and let $\Omega_0$ be a free parameter in the last two columns.}
\label{table:CCE_parameter_estimation}
\end{table}

These results underscore several key advantages of using BOB for data analysis. Its high accuracy near the waveform peak circumvents the difficult problem of choosing a ringdown start time, as required by overtone models. Because it can be used with no or minimal coefficient fitting, it provides a more robust method for estimating remnant parameters from noisy detector data. Furthermore, its analytic link between the peak amplitude (i.~e.~the loudest part of the signal) and the fundamental mode provides an independent and powerful cross-check on QNM parameters extracted from the quieter, often noise-dominated ringdown.

\subsection{IMR comparisons}
While the most natural comparisons to BOB are overtone models, BOB can easily be used in full inspiral-merger-ringdown (IMR) waveform models, as is done in \cite{Mahesh:2025}, due to its high accuracy all the way back to the peak of the waveform. Therefore, it is useful to assess the accuracy of BOB against the merger-ringdown portion of IMR models as well. We compare BOB to SEOBNRv5HM \cite{SEOBNR_cite2}, a highly-calibrated Effective-One-Body model, and NRSurHyb3dq8 \cite{surr_cite1}, an NR surrogate model. An overview of the NR parameter space used to calibrate SEOBNRv5HM can be found in Fig.~2 of \cite{SEOBNR_cite2}.

\begin{figure}
    \centering
    \includegraphics[width=\linewidth]{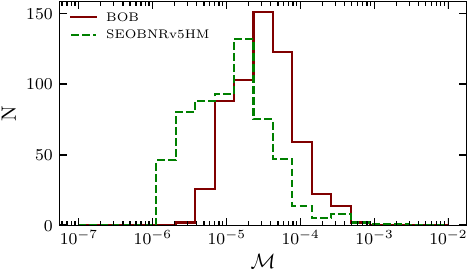}
    \caption{Comparison between BOB (maroon) and SEOBNRv5HM (dashed green) for the $\mathcal{N}_{22}$ mismatch in the time interval $[t_p^{\mathcal{N}_{22}},t_p^{\mathcal{N}_{22}}+75M]$ across all quasi-circular and non-precessing cases in the SXS catalog.}
    \label{fig:news_mismatch_all}
\end{figure}

\begin{figure}
    \centering
    \includegraphics[width=\linewidth]{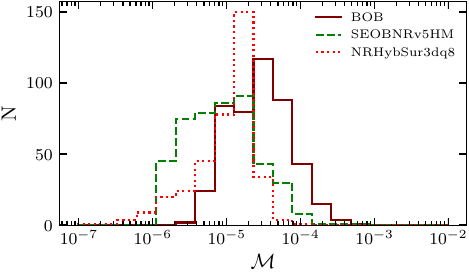}
    \caption{Comparison of \news mismatch in the time interval $[t_p^{\mathcal{N}_{22}},t_p^{\mathcal{N}_{22}}+75M]$ across all quasi-circular and non-precessing cases in the SXS catalog with $q \leq 8$ and $\chi_i \leq 0.8$, consistent with the training limits of NRHybSur3dq8.}
    \label{fig:news_mismatch_surr_lim}
\end{figure}

\begin{figure*}
    \includegraphics[]{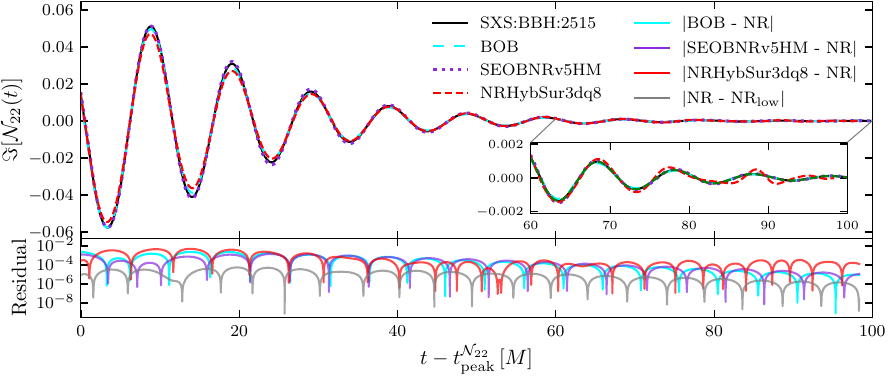}
    \caption{Case with the worst $\mathcal{N}_{22}$ mismatch for BOB. Because this case is outside the domain of all NR surrogate models, NRHybSur3dq8 is forced to extrapolate beyond its training domain. In the bottom panel we plot the residual for the imaginary part of \news\,between NR and BOB, between NR and SEOBNRv5HM, between NR and NRHybSur3dq8, and between the highest (Lev4) and second-highest (Lev3) resolution NR waveform. SXS:BBH:2515 \cite{SXS:BBH:2515,SXSCatalogData_3.0.0} consists of initial spins $\vec{\chi}_1\cdot \hat{z} = 0.8$, $\vec{\chi}_2\cdot \hat{z} = 0.5$, mass ratio $q = 10$, $M_f = 0.981$ and $\vec{\chi}_f \cdot \hat{z} = 0.855$. Because the mismatch as defined in Eq.~(\ref{eq:mismatch}) is independent of an overall amplitude normalization, we rescale all waveforms to a common peak amplitude.} \label{fig:worst_BOB}
    \includegraphics[]{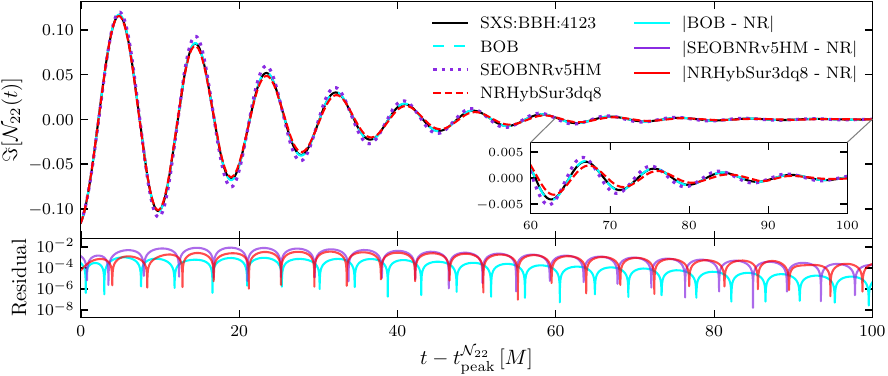}
    \caption{Case with the worst $\mathcal{N}_{22}$ mismatch for SEOBNRv5HM. Because this case is outside the domain of all NR surrogate models, NRHybSur3dq8 is forced to extrapolate beyond its training domain. To eliminate any potential bias in BOB from the use of NR remnant parameters, in this figure we use the remnant parameters obtained from the same fits used by SEOBNRv5HM for BOB. In the bottom panel we plot the residual for the imaginary part of \news\, between NR and BOB, NR and NRHybSur3dq8, as well as for NR and SEOBNRv5HM. For this case only one resolution level (Lev4) is available for the NR data. SXS:BBH:4123 \cite{SXS:BBH:4123,SXSCatalogData_3.0.0} consists of initial spins $\vec{\chi}_1\cdot \hat{z} = 0.9$, $\vec{\chi}_2\cdot \hat{z} = -0.5$, mass ratio $q = 4$, $M_f = 0.952$ and $\vec{\chi}_f \cdot \hat{z} = 0.915$. Because the mismatch as defined in Eq.~(\ref{eq:mismatch}) is independent of an overall amplitude normalization, we rescale all waveforms to a common peak amplitude.}
    \label{fig:worst_EOB}
\end{figure*}

We begin by assessing the accuracy of \news, the fundamental quantity BOB models most accurately. Since the IMR models produce strain, we compute the IMR \news\, via a cubic spline differentiation. As shown in Figs.~\ref{fig:news_mismatch_all} and (\ref{fig:news_mismatch_surr_lim}), across the full non-precessing parameter space, BOB's median $\mathcal{N}_{22}$ mismatch of $3.18 \times 10^{-5}$ is close to SEOBNRv5HM's $1.18 \times 10^{-5}$. Within the training domain of the surrogate, BOB's performance of $2.88\times10^{-5}$ is again competitive with both SEOBNRv5HM ($1.31\times10^{-5}$) and NRSurHyb3dq8 ($9.26 \times10^{-6}$). This demonstrates that BOB, a largely uncalibrated model, captures the merger-ringdown radiation with accuracy approaching that of NR reliant frameworks. In Figs.~\ref{fig:worst_BOB} and (\ref{fig:worst_EOB}), we show the cases that produce the worst mismatch with BOB and SEOBNRv5HM. Both of these cases involve configurations outside the domain covered by any NR surrogate model. The worst mismatch case for SEOBNRv5HM corresponds to a highly spinning unequal mass system ($q=4, \, \chi_1 = 0.9$) which represents an area of the parameter space not well covered by the SXS catalog. This causes a significant drop in accuracy for the NR-reliant merger ringdown model in SEOBNRv5HM. BOB's accuracy, in contrast, remains robust. 

\begin{figure*}
    \centering
    \includegraphics[]{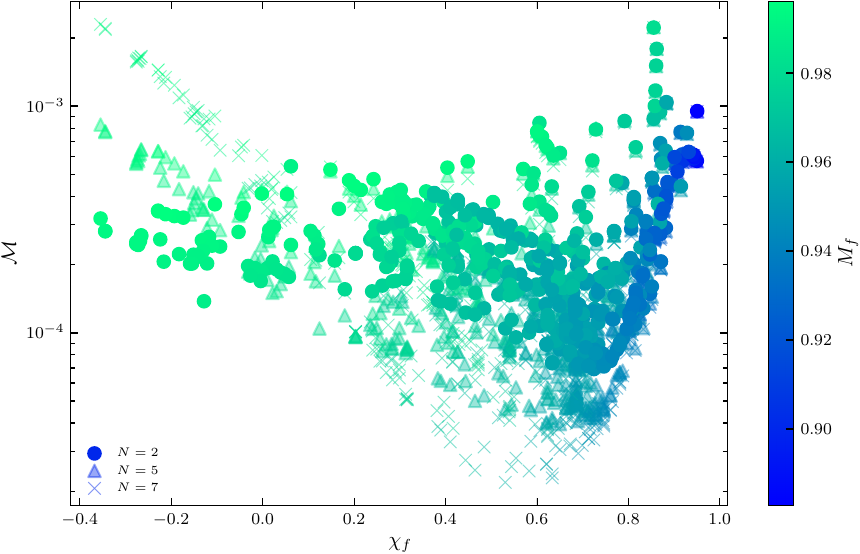}
    \caption{$h_{22}$ mismatch comparison of BOB against NR simulations taking three different values for $N$ in Eq.~\eqref{eq:strain_from_news_series}.}
    \label{fig:strain_various_N}
\end{figure*}

While we have shown that BOB fundamentally models \news, IMR models need to produce \strain. However, this conversion is not trivial and we are unable to find a closed form solution for the integration of BOB. To maintain a fully analytic framework, we develop an approximate solution using an asymptotic series, detailed in Sec.~\ref{sec:strain_from_BOB}. This approach requires taking high-order derivatives, and we find that numerical derivatives are not sufficiently accurate. Therefore we utilize auto-differentiation with \texttt{JAX} \cite{jax2018github} to obtain accurate high-order derivatives. This highlights another benefit of BOB's analytic and smooth nature; it makes BOB easily adaptable to gravitational wave codes that utilize the auto-differentiation and GPU friendly benefits of \texttt{JAX}. While in principle computer algebra systems like $\texttt{SymPy}$ \cite{SymPy} can also be used, our analytical approximation for strain can become extremely large for sufficiently high N, which can lead to practical difficulties, such as correctly simplifying the equations, with \texttt{SymPy}. Because Eq.~\eqref{eq:strain_from_news_series} is a divergent series, there is an optimal number of terms, $N$, that provides the best approximation \cite{boyd1999devil}. In Table.~\ref{table:strain_series_N} we list the median and mean values we obtain for each choice of $N$. However, as shown in Fig.~\ref{fig:strain_various_N}, we notice that while for nearly all systems $N=7$ provides the highest accuracy strain, for systems with the remnant spin pointing opposite the direction of the initial orbital angular momentum, $N=2$ provides higher accuracy results. The exact reason for this is unclear, but as argued in the previous section, these cases will likely have the highest impact from retrograde modes not currently modeled by BOB.

\begin{table}[htb]
\begin{ruledtabular}
\begin{tabular}{ccc}
\(N\) & Mean & Median \\ \hline
0 & \(5.37 \times 10^{-4}\) & \(5.12 \times 10^{-4}\) \\
1 & \(3.00 \times 10^{-4}\) & \(2.71 \times 10^{-4}\) \\
2 & \(2.64 \times 10^{-4}\) & \(2.15 \times 10^{-4}\) \\
3 & \(2.36 \times 10^{-4}\) & \(1.74 \times 10^{-4}\) \\
4 & \(2.40 \times 10^{-4}\) & \(1.75 \times 10^{-4}\) \\
5 & \(2.23 \times 10^{-4}\) & \(1.41 \times 10^{-4}\) \\
6 & \(2.45 \times 10^{-4}\) & \(1.55 \times 10^{-4}\) \\
7 & \(2.58 \times 10^{-4}\) & \(1.27 \times 10^{-4}\) \\
8 & \(4.01 \times 10^{-4}\) & \(1.63 \times 10^{-4}\) \\
\end{tabular}
\end{ruledtabular}

\caption{Mean and median values for the $h_{22}$ mismatch against the quasi-circular and non-precessing SXS catalog for the number of terms $N$ included in the summation in Eq.~\eqref{eq:strain_from_news_series}.}
\label{table:strain_series_N}
\end{table}

\begin{figure}
    \centering
    \includegraphics[]{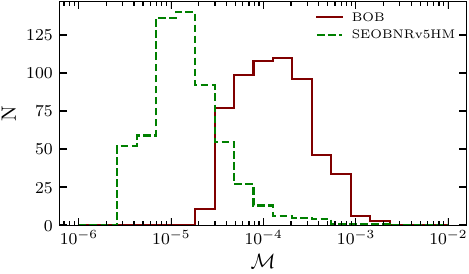}
    \caption{Comparison of the $h_{22}$ mismatch in the time interval $[t_p^{h_{22}},t_p^{h_{22}}+75M]$ across all quasi-circular and non-precessing cases in the SXS catalog between BOB (maroon) and SEOBNRv5HM (dashed green).}
    \label{fig:Mismatch_omega0_optimized_varied_N}
\end{figure}

\begin{figure}
    \centering
    \includegraphics[]{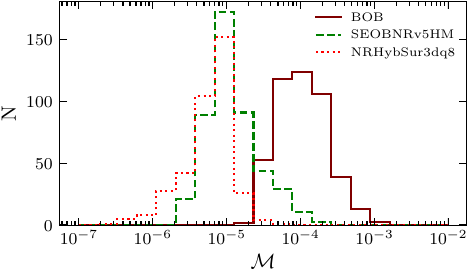}
    \caption{Comparison of the $h_{22}$ mismatch in the time interval $[t_p^{h_{22}},t_p^{h_{22}}+75M]$ across all quasi-circular and non-precessing cases in the SXS catalog with $q \leq 8$ and $\chi_i \leq 0.8$, consistent with the training limits of NRHybSur3dq8, between BOB (maroon), SEOBNRv5HM (dashed green), and NRHybSur3dq8 (dotted red).}
    \label{fig:Mismatch_omega0_optimized_surr_varied_N}
\end{figure}

Finally, we compare the mismatch of \strain\, between our analytic BOB waveform and IMR models. The results are shown in Figs.~\ref{fig:Mismatch_omega0_optimized_varied_N} and \ref{fig:Mismatch_omega0_optimized_surr_varied_N}. We observe a significant increase in BOB's mismatch relative to the IMR models, particularly near the waveform peak. This is not an inherent deficiency of the BOB model's core physics, but a direct consequence of our standalone approach to integrating \news. Our analytic news to strain conversion lacks information from the inspiral, such as the peak strain amplitude, which IMR models use to seamlessly stitch the inspiral and merger waveforms. This discrepancy is most pronounced near the peak. As Fig.~\ref{fig:mismatch_p10} shows, if we begin our comparison just 10M after the peak, BOB's strain mismatch becomes comparable to that of the EOB model. This confirms that the main source of error is in the attachment region.

\begin{figure}
    \centering
    \includegraphics[]{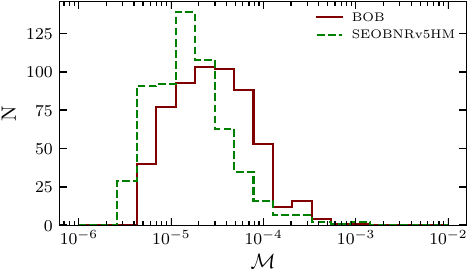}
    \caption{Comparison of the $h_{22}$ mismatch in the time interval $[t_p^{h_{22}}+10M,t_p^{h_{22}}+75M]$ across all quasi-circular and non-precessing cases in the SXS catalog between BOB (maroon) and SEOBNRv5HM (dashed green).}
    \label{fig:mismatch_[10,75]}
\end{figure}
\begin{figure}
    \centering
    \includegraphics[]{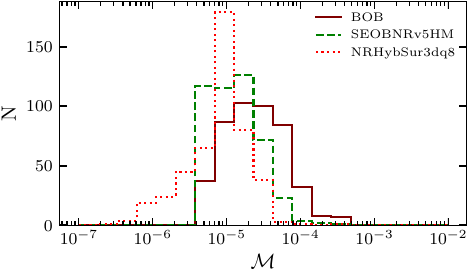}
    \caption{Comparison of the $h_{22}$ mismatch in the time interval $[t_p^{h_{22}}+10M,t_p^{h_{22}}+75M]$ across all quasi-circular and non-precessing cases in the SXS catalog with $q \leq 8$ and $\chi_i \leq 0.8$, consistent with the training limits of NRHybSur3dq8, between BOB(maroon), SEOBNRv5HM (dashed green), and NRHybSur3dq8 (dotted red).}
    \label{fig:mismatch_p10}
\end{figure}

\subsection{Probing the limits of NR}
\begin{figure}
    \centering
    \includegraphics[]{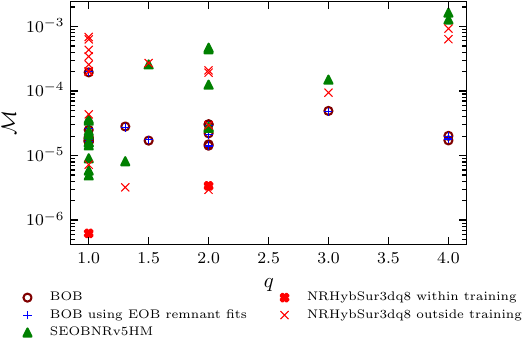}
    \caption{Comparison of the $\mathcal{N}_{22}$ mismatch between NR and BOB (maroon circles) and NR and SEOBNRv5HM (green triangles) for cases with $\chi_f \geq 0.9$ as a function of the initial mass ratio $q$. To ensure that there is no bias due to the initialization of BOB with NR remnant parameters, we also plot the mismatches (blue plus) between NR and BOB initialized with remnant parameters using the same NR fits as in SEOBNRv5HM. Lastly, we plot the mismatches between NR and NRHybSur3dq8. Because many of these simulations are outside the training regime of NRHybSur3dq8, we separately plot the two cases within the training regime (thick red x) and the 18 cases outside the training regime (thin red x).}
    \label{fig:news_mismatch_edge}
\end{figure}

While merger-ringdown approaches that rely heavily on NR can obtain high levels of accuracy, they are inherently tied down to the size and accuracy of NR catalogs. Future detectors will demand not only higher accuracy from NR catalogs, but also broader coverage of a high-dimensional parameter space. The current largest NR catalog  is estimated to have required 480,000,000 core hours of computation time \cite{sxs_cat3}. This highlights a critical advantage of the minimally tuned yet highly accurate approach of BOB - \textit{its accuracy is not fundamentally limited by the available NR data}. We demonstrate this advantage in Fig.~\ref{fig:news_mismatch_edge}, where we examine systems with high remnant spin, $\chi_f \geq 0.9$, a region of the parameter space that is notoriously difficult for NR. No non-precessing system with initial mass ratio $q>4$ and remnant spin $\chi_f \geq 0.9$ currently exists in the public SXS catalog. The vast majority of the cases shown in Fig.~\ref{fig:news_mismatch_edge} fall outside the domain of validity for surrogate models, therefore for the mismatches between NRHybSur3dq8 and NR, we separately plot the cases within and outside the training space. All cases considered in Fig.~\ref{fig:news_mismatch_edge} have lower mismatches between the highest and second highest NR resolution levels than the model mismatch against NR. In this regime, we observe a considerable drop in the accuracy of the highly-calibrated SEOBNRv5HM model as the mass ratio increases beyond unity.\footnote{In Fig.~\ref{fig:news_mismatch_edge}, only one case for $q=2$ has a lower mismatch for SEOBNRv5 ($2.81\times10^{-5}$) than BOB ($3.03 \times 10^{-5}$)}. Likewise, NRHybSur3dq8 struggles to extrapolate accurately to this corner of the parameter space. BOB, on the other hand, shows performance consistent with its accuracy across the rest of the parameter space due to its minimal reliance on NR catalogs. This makes BOB a powerful tool for independently validating NR-based models and for quality control of NR catalogs, particularly at the edges of their coverage. Furthermore, future analytical improvements to the BOB formalism will enhance its accuracy globally and independently of improvements to NR catalogs.

\section{Superkicks}
\label{sec:superkicks}
\begin{figure}
    \centering
    \includegraphics[]{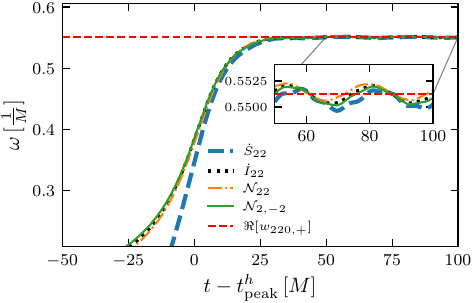}
    \caption{Frequency of $\mathcal{N}_{22}$, $\mathcal{N}_{2,-2}$, $\dot{I}_{22}$ and $\dot{S}_{22}$ for the q1\_superkick configuration available in the public EXT-CCE database.}
    \label{fig:current_frequency}
\end{figure}

A previous study of BOB in the context of superkicks \cite{universal_superkick} found BOB unable to accurately model the current quadrupole wave or any of its derivatives. The current and mass quadrupole waves are obtained from combining the $h_{lm}$ and $h_{l,-m}$ waveforms as
\begin{align}
    I_{lm} = \frac{1}{\sqrt{2}}\bigg[h_{lm} + (-1)^mh^*_{l,-m}\bigg]\\
    S_{lm} = \frac{1}{\sqrt{2}}\bigg[h_{lm} - (-1)^mh^*_{l,-m}\bigg]
\end{align}
Because BOB best models \news, we focus on the derivatives of the quadrupole waves $\dot{I}_{lm}, \,\dot{S}_{lm}$. The authors of \cite{universal_superkick} treated BOB as a more phenomenological model depending on six free parameters, and it is unclear whether the quadrupole waves were obtained by creating the appropriate $(l,m)$ modes with BOB or using the BOB amplitude and frequency evolution equations to directly model $\dot{I}_{22}$ and $\dot{S}_{22}$. Here we show that by using BOB to model $\mathcal{N}_{22}$ and $\mathcal{N}_{2,-2}$ and constructing $\dot{I}_{22}$ and $\dot{S}_{22}$ from these two waveforms, we can accurately model the quadrupole waves. Since the BOB frequency evolution in Eq.~\eqref{eq:BOB_news_frequency_evolution} is based on the assumption that Eq.~\eqref{eq:BOB_amplitude} is modeling \news, this is the proper way to construct the quadrupole quantities. Furthermore, we do not allow $A_p$ and $t_p$ to be free parameters that are obtained through a fit, but rather set them to the peak amplitude of the waveform and the corresponding time. In Fig.~\ref{fig:current_frequency}, we show that while the mass quadrupole frequency exhibits a similar structure to the $(2,2)$ and $(2,-2)$ modes, the current quadrupole frequency clearly differs. If the mass and current quadrupole waves were constructed directly, using the BOB equations as a phenomenological ansatz, this may explain why the current quadrupole was unable to be successfully modeled in \cite{universal_superkick}, whereas our construction of BOB models the current quadrupole quite accurately. 

In Table \ref{Table:superkick_values}, we use the superrest transformed q1\_superkick configuration available in the public EXT-CCE database to evaluate BOB's ability to model the mass and current quadrupole wave \footnote{This simulation is different from the one considered in \cite{universal_superkick}, but we find consistent results for all superkick simulations.}. For consistency with \cite{universal_superkick} we determine our values from the peak time of the $L^2$ norm of \strain\, to 100M afterwards. Unlike in \cite{universal_superkick}, we do not find any significant bias in BOB's construction of the current quadrupole. Similar to our earlier results, we find that the simple error (Eq.~\ref{eq:simple_erorr}) can show significant sensitivity based on our fitting choices. To highlight this for $\mathcal{N}_{22}$ and $\mathcal{N}_{2,-2}$ in parentheses we also provide the values obtained through an alternative fitting approach. Rather than have the optimization algorithm search over $M_f,\chi_f$ and $\Omega_0$, which is the approach used for the entries in Table \ref{Table:superkick_values} not inside parentheses, for the parenthetical values we first have the same algorithm search only over $M_f$ and $\chi_f$, then we search for the value of $\Omega_0$ that minimizes the least-squares difference between the BOB frequency evolution and the NR data. As seen in Table \ref{Table:superkick_values}, both approaches yield similar mismatch values but can differ by a factor of two in the accuracy of the mass and three in the accuracy of the spin. In particular, as seen in the $\mathcal{N}_{2,-2}$ column, a lower mismatch does not correlate with a more accurate recovery of parameters. Furthermore, once again, we find that the most accurate way to recover parameters is to isolate the late ringdown section of the waveform, where our two fitting approaches yield more consistent results and significant increases in the accuracy of the recovered parameters. 

Outside of the current quadrupole, our results are largely consistent with those of \cite{universal_superkick} for $\mathcal{N}_{2,\pm2}$ and $\dot{I}_{22}$. While we are unable to match the accuracy of using 8 QNMs, the usage of 8 QNMs requires 16 parameters that must be fit to the waveform. In our comparisons here, we fit, at most, one parameter to the waveform: $\Omega_0$. The fairest comparison to BOB would be using the same number of free parameters in a QNM based model. In such a case, BOB vastly outperforms the QNM model. In general, we find it takes 4--8 QNMs to match BOB's accuracy for the same set of SXS superkick configurations studied in \cite{universal_superkick}. Lastly, we must emphasize that our frequency evolution for \strain\, is significantly different than the one presented in \cite{universal_superkick}; in particular, the analytic frequency evolution derived in Appendix B2 of \cite{universal_superkick} does not asymptote to the QNM frequency, unlike their frequency evolution for \psi4 and \news, but rather asymptotes to $\Omega_0A_p\sqrt{2\tau}$ as $t \rightarrow \infty$.

\begin{table*}[]
\centering
\renewcommand{\arraystretch}{1.5}
\begin{tabular}{lcccc}
\toprule \hline
&$\mathcal{N}_{(2,2)}$ & $\mathcal{N}_{(2,-2)}$ & $\mathcal{\dot{I}}_{(2,2)}$ & $\mathcal{\dot{S}}_{(2,2)}$ \\
\midrule 
$\Delta M/M_f \,[t^h_p,t^h_p{+}100M]$ ($\times 10^{-3}$) & 7.2 (13.0) & 11.9 (5.98) & 6.12 & 8.1 \\
$\Delta \chi \, [t^h_p,t^h_p{+}100M]$ ($\times 10^{-3}$) & 5.6 (14.0) & 14.72 (5.6) & 6.93 & 14.7 \\
$\epsilon \, [t^h_p,t^h_p{+}100M]$ ($\times 10^{-3}$) & 9.2 (19.1) & 18.93 (8.2) & 9.24 & 16.73 \\
$\mathcal{M}\, [t^h_p,t^h_p{+}100M]$ ($\times 10^{-5}$) & 2.62 (3.98) & 4.32 (5.84) & 3.51 & 3.85 \\\hline
$\epsilon \, [t^h_p+40M,t^h_p{+}100M]$ ($\times 10^{-3}$) & 1.09 (0.99) & 0.84 (0.92) & 1.01 & 1.43 \\
$\mathcal{M}\, [t^h_p+40M,t^h_p{+}100M]$ ($\times 10^{-5}$) & 0.14 (0.15) & 0.16 (0.16) & 0.15 & 0.16 \\
\bottomrule
\end{tabular}
\caption{Mismatch and parameter recovery error, Eq.~\eqref{eq:simple_erorr}, for the superrest transformed q1\_superkick configuration available in the public EXT-CCE database. $\Omega_0$ in BOB is a free parameter in the optimization algorithm. In parentheses, we provide the values obtained from a second fitting approach where $\Omega_0$ is obtained from a least squares fit to the NR data.}
\label{Table:superkick_values}
\end{table*}

\section{Discussion}
\label{sec: discussion}
In this work, we performed a comprehensive analysis of the Backwards-One-Body (BOB) model for quasi-circular and non-precessing systems. We demonstrated that BOB, a model based on the divergence of null geodesics from the remnant's light ring, provides a high accuracy description of the merger-ringdown radiation. Our analysis confirms that BOB most accurately models the gravitational wave news, $\mathcal{N}$, achieving accuracies comparable to state-of-the-art EOB and surrogate based models while remaining minimally reliant on NR catalogs. The model depends only on the remnant properties ($M_f,\chi_f$) and an initial frequency parameter, $\Omega_0$. In this work, we have intialized BOB with the NR remnant properties while SEOBNRv5HM has been initialized with the NR initial parameters. SEOBNRv5HM then uses a highly accurate NR fit to obtain the remnant properties \cite{SEOBNR_cite2}. In Figs.~\ref{fig:worst_EOB} and (\ref{fig:news_mismatch_edge}) we use the remnant values determined from the same NR fits used in SEOBNRv5HM for BOB and find no significant increase in mismatch compared to using the remnant values obtained directly from the NR simulation. We note that the mismatch, as defined in Eq.~\ref{eq:mismatch}, is independent of the peak amplitude value, $A_p$. Thus our mismatch values for BOB are not biased by using the NR data for $A_p$.

Focusing on \news, we compared BOB both qualitatively and quantitatively to overtone-based models. We showed how BOB's analytic structure naturally incorporates information from an infinite series of overtones and that the ``knee'' seen in the accumulated mismatch, similar to the ``knee'' in QNM mismatches, generally coincides with the estimated onset of stable QNMs. Across the parameter space, BOB's accuracy is comparable to a 4 to 8 QNM model, but it achieves this without the need to fit the 8 to 16 free parameters required by such models.
A unique prediction of BOB is the analytic relationship between the peak news amplitude and the amplitude of the fundamental QNM. We verified this prediction against CCE waveforms and found it to be accurate to within 1--2\%. While perturbation theory suggests that the source term describing the perturbation should contain a dependence on the overtone index $n$, a number of studies \cite{cheung2024extracting,improved_qnm_extraction,importance_of_overtones,nonlinear_qnm} have found that the perturbation to the remnant seems to be largely independent of $n$. This finding, which has required extensive numerical relativity simulations, is naturally found in BOB's approach to merger ringdown modeling.

Because BOB is minimally reliant on NR, future improvements to its underlying physics will enhance its accuracy globally. By linking a drop in BOB's accuracy to the increasing importance of retrograde modes in systems with $\chi_\mathrm{eff}<0$, we identify a significant example where adding additional physics into BOB can improve its accuracy. Since the BOB formalism can be trivially extended to include these modes, we anticipate that this addition could improve the model's accuracy by an order of magnitude for these systems. We leave this for future work, which will benefit greatly from an expanded public CCE waveform database.

Our work also reveals several opportunities where BOB can further enhance current ringdown analysis efforts. Given its analytic, noise-free nature, and its rich overtone-like structure, BOB provides an ideal testbed for assessing the robustness of QNM fitting algorithms. Furthermore, the residual between BOB and NR waveforms contains clear signatures of unmodeled physics, suggesting BOB could be a powerful tool for isolating and identifying non-linear modes. For data analysis, it offers a robust method for obtaining inspiral-independent remnant properties from noisy detector data, and can provide independent tests of overtone model based results. Lastly, BOB is highly suitable for codes using \texttt{JAX}, whose auto-differentiation capabilities provide a natural synergy with BOB's analytic form.

We also compare the accuracy of BOB to the Effective One Body based SEOBNRv5HM and the NR surrogate based NRHybSur3dq8. We show BOB can model the merger-ringdown \news\, radiation with comparable accuracy. However, IMR models need to obtain \strain, not \news. We use an analytic approximation to convert the BOB \news\, into \strain, but we do find a larger mismatch compared to IMR models. This discrepancy is not an inherent deficiency of the model's physics, but a direct consequence of performing the conversion without information from the inspiral, such as the peak strain amplitude. As we showed, this mismatch is largest near the peak and becomes comparable to IMR models just 10M later.

Current waveform modeling is heavily dependent on computationally expensive NR simulations. The SXS catalog alone is estimated to have required 480 million core hours \cite{sxs_cat3}. To meet the needs of future ground and space based detectors, both numerical relativity waveforms and IMR waveform models must see significant improvements in their accuracy \cite{jan2024accuracy,hu2022assessing,ferguson2021assessing}, which will require a considerable amount of additional computational resources. Our work demonstrates that BOB, a minimally tuned and physically motivated merger-ringdown model, can match the accuracy of NR based approaches. Because BOB's accuracy is not tethered to the size and accuracy of NR catalogs, it retains its accuracy in regions of the parameter space where NR simulations are sparse. Therefore, BOB can provide independent tests of NR based waveform models and be useful in quality control for NR catalogs. Furthermore, improvements to the analytical sector of BOB will globally increase its accuracy, independently of improvements to NR catalogs.

While this work has focused on the $(2,2)$ mode of quasi-circular and non-precessing systems, BOB is easily adaptable to higher modes and precessing systems. In the future, we will provide a rigorous assessment of BOB's accuracy throughout the entire parameter space. We will also further investigate how BOB can illuminate the fundamental physics of the merger, such as the identification of non-linear modes and the physical onset of the linear ringdown. Lastly, we also provide a python package, \texttt{gwBOB} \cite{gwBOB}, that simplifies the process of constructing various flavors of BOB, allowing for the easy usage of BOB merger-ringdown waveforms in a variety of research problems.

\begin{acknowledgments}
AK and STM were supported in part by NSF CAREER grant PHY-1945130 and NASA grants 22-LPS22-0022 and 24-2024EPSCoR-0010. This research was made possible by the NASA West Virginia Space Grant Consortium, Grant \# 80NSSC20M0055.  The authors acknowledge the computational resources provided by the WVU Research Computing Thorny Flat HPC cluster, which is funded in part by NSF OAC-1726534. This work, and the associated python package, used the following software packages: \textsc{gwBOB} \cite{gwBOB}, \textsc{pyseobnr} \cite{pyseobnr}, \textsc{gwsurrogate} \cite{gwsurrogate}, \textsc{scri} \cite{scri1,scri2,scri3,scri4}, \textsc{sxs} \cite{Boyle2025SXS}, \textsc{qnm} \cite{Stein:2019mop}, \textsc{surfinBH} \cite{surfinBH,surr_ringdown,varma_stein_gerosa_2018_surfinBH}, \textsc{qnmfinder} \cite{improved_qnm_extraction}, \textsc{qnmfits} \cite{qnmfits,qnmfits2}, \textsc{kuibit} \cite{kuibit}, \textsc{JAX} \cite{jax2018github}, \textsc{numpy} \cite{numpy}, \textsc{scipy} \cite{scipy} and \textsc{matplotlib} \cite{matplotlib}.
\end{acknowledgments}

\bibliography{citations}

\end{document}